\documentclass[A4paper,11pt]{article}
\usepackage{kotex}
\usepackage{graphics,graphicx}
\usepackage{epsfig,wrapfig,color}
\usepackage[ruled,vlined]{algorithm2e}
\usepackage{longtable}
\usepackage{amsmath,amssymb}
\usepackage{verbatim,,lscape}
\usepackage{natbib}
\usepackage{lastpage}
\usepackage{multirow}
\usepackage{url}
\newcommand{\hide}[1]{}
\usepackage[normalem]{ulem}
\usepackage{bm}
\usepackage{authblk}
\usepackage[unicode,colorlinks,citecolor=blue]{hyperref}

\makeatletter

\usepackage[outerbars,color]{changebar}
\ifx\pdfoutput\undefined
\else\ifnum\pdfoutput>0
  \usepackage{pdfcolmk}
\fi\fi
\cbcolor{black}

\usepackage[margin=1.0in]{geometry}
\usepackage{tikz}
\usetikzlibrary{bayesnet}
\usetikzlibrary{fit,positioning}

\newfont{\rmm}{cmr10 at 11pt}
\rmm

\usepackage[english]{babel}

\def\BibTeX{{\rm B\kern-.05em{\sc i\kern-.025em b}\kern-.08em
		T\kern-.1667em\lower.7ex\hbox{E}\kern-.125emX}}

\begin{document}
\title{Quantile Regression with Multiple Proxy Variables}
\author[1]{Dongyoung Go}
\author[1,2]{Ick Hoon Jin}
\author[1,2]{Jongho Im}
\affil[1]{Department of Statistics and Data Science, Yonsei University}
\affil[2]{Department of Applied Statistics, Yonsei University}
\date{}

\maketitle

\begin {abstract}
Data integration has become increasingly popular owing to the availability of multiple data sources. This study considered quantile regression estimation when a key covariate had multiple proxies across several datasets. In a unified estimation procedure, the proposed method incorporates multiple proxies that have various relationships with the unobserved covariates. The proposed approach allows the inference of both the quantile function and unobserved covariates. Moreover, it does not require the quantile function’s linearity and, simultaneously, accommodates both the linear and nonlinear proxies. Simulation studies have demonstrated that this methodology successfully integrates multiple proxies and revealed quantile relationships for a wide range of nonlinear data. The proposed method is applied to administrative data obtained from the Survey of Household Finances and Living Conditions provided by Statistics Korea, to specify the relationship between assets and salary income in the presence of multiple income records.
\end{abstract}
 
\clearpage 

\section{Introduction}\label{intro} 

With the emergence of  multiple data sources, such as administrative, web-collected, and big data, data integration has become popular. For example, the integration of administrative and statistical data  has been widely used in official statistics  to improve data quality, data analyses, and data collection costs \citep{berg2021}. As the demand for data integration increases, two kinds of statistical methodologies have been proposed---namely, record linkage and statistical matching. Record linkage assumes overlapped units, which allow the linkage of the same units between data sources, while statistical matching frequently links similar units between non-overlapped datasets \citep{leulescu13}. In practice, record linkage and statistical matching are often applied separately depending on whether the same unit is matched. In this study, we are interested in  combining multiple values observed for the same attribute, a common issue in record linkage. 

When multiple data sources are available, it is easy to have multiple observations on the same attribute  due to nonresponse or measurement errors, different observation times, and mode effects. To handle these multiple observations in statistical models, we treat them as multiple proxies of the unobserved true value. These proxies include various domains, from the case in which covariates are measured with errors \citep{carroll2006measurement, fuller2006measurement} when the true covariate is simply unobserved \citep{filmer2001estimating}, to when the variable of interest is a conceptual variable that is difficult to examine precisely  \citep{solon1992intergenerational, zimmerman1992regression, mazumder2001earnings}. Specifically, we consider the problem of estimating the quantile function in the presence of multiple proxies for true covariates. As in conventional linear regression, the quantile regression estimator’s inconsistency in the absence of a true covariate is a commonly discussed issue in the literature \citep{brown1982robust, he2000quantile, carroll2006measurement, wei2009quantile, montes2011quantile, hausman2021errors}. 

Several studies have incorporated proxy variables in quantile regression estimation. \cite{he2000quantile}---considering the problem of estimating quantile regression coefficients in errors-in-variables models with a proxy variable---proposed  an estimator in the context of linear and partially linear models. \cite{wei2009quantile} presented a nonparametric method for correcting bias caused by measurement error in the linear quantile regression model by constructing joint estimating equations that simultaneously hold for all quantile levels. Further, \cite{firpo2017measurement} proposed a semiparametric two-step estimator when repeated measures for the proxy are available. \cite{schennach2007instrumental, schennach2008quantile} discussed identifying a nonparametric quantile function under various settings when an instrumental variable is measured on all sampling units. \cite{wang2012corrected} modified the standard quantile regression objective function, tailoring it to the Gaussian measurement error model. 

However, these studies are limited to combining multiple proxies in a non-unified framework and frequently require repeated observations of proxy variables. To address these issues, we propose a novel Bayesian measurement error model by 
specifying how the proxies are generated from unobserved covariates. Motivated by the measurement error model approach \citep{clayton1992, richardson1993, lubotsky2006interpretation, fuller2006measurement}, we demonstrate a Bayesian flexible combining method that can (i) integrate information from multiple proxies related to unobserved covariates in a wide range of different formulas and (ii) simultaneously deal with the flexible quantile regression model with an estimation of the unobserved covariate and the unobserved relationship to the proxy. Specifically, we first consider a generalized additive model for the proxies such that these are decomposed as a smoothing function of the true covariate and unobserved additive error; by contrast, most studies have employed a classical measurement error model wherein a zero-mean error is simply added to the unobserved covariate. Moreover, we employ a factor analysis-based  measurement error model introduced in \citet{fuller2006measurement} to combine multiple proxies in a unified framework. We treat the true covariate as a common factor in the factor analysis and, thereafter, construct a set of additive models that include a smoothing function of the same true covariate. This approach allows the generation of synthetic values using a Bayesian approach \citep{clayton1992, richardson1993}. However, following \cite{berry2002bayesian}, we use the Bayesian nonparametric quantile regression function as the outcome model with a natural cubic spline \citep{thompson2010bayesian} and penalized spline \citep{lang2004bayesian}, rather than linear parametric models that might be sensitive to model misspecification. 

The proposed method has several advantages. First, we incorporate a more flexible form of proxy variables, previously limited to additive measurement error, thereby enabling us to account for proxies with both linear and nonlinear relationships with a true covariate. Second, the proposed method incorporates an arbitrary number of multiple proxies and infers true covariate and quantile regression functions. Third, the proposed framework does not assume the linearity of the quantile regression function, which restricts the model’s flexibility. 

The remainder of the paper is organized as follows: In section \ref{setup}, we describe the basic setup of the investigation. In section \ref{our_method}, we introduce a method of combining multiple proxies and making inferences regarding nonparametric quantile regression likelihood. Further, the related prior and detailed Gibbs sampling steps are described. In section \ref{simulation}, we present the simulation results for various simulation data. Extensive simulation studies on various datasets reveal the approach’s effectiveness in incorporating multiple proxies simultaneously, compared with the method of using one proxy variable directly and the previously proposed structural model. In section \ref{real_data}, we apply the proposed method to the public administrative dataset, which includes various economic features of 18,064 families, such as salary and property income. We apply this methodology to administrative data to study the quantile relationship between assets and actual salary income. In section \ref{conclude}, we provide concluding remarks .

\section{Basic Setup}\label{setup}

Let $\{y_i,x_i\}^n_{i=1}$ be a random sample of size $n$, where $y_i$ is the outcome variable and $x_i$ is the explanatory covariate. Let $g_p(x_i)$ be the $p$-th quantile of the conditional distribution of $y_i$ considering $x_i$ such that 
\begin{equation}
	P\Big(y_i \leq g_p(x_i)\Big)=p, \quad 0<p<1
	\addtocounter{equation}{1}\tag{\theequation}, \label{QR}
\end{equation}
Suppose that covariate $x_i$ is not directly observed, but multiple proxies related to the covariate are observed from multiple data sources. These proxies are denoted as $w_{ki}$, $k=1,\ldots,K$. We employ popular methods (e.g., \citealp{li1998nonparametric, carroll2006measurement, delaigle2008deconvolution}) to estimate the quantile function $g_p(x_i)$ when proxies are replicates of mismeasured variables, wherein only the mean zero measurement error is added to the covariate. However, these methods cannot be directly extended when multiple proxies exist in a different relationship to the unobserved $x$.

To simultaneously account for different proxies, first, we write a set of regression models such that 
\begin{align}
y_{i} & =g_p(x_{i})+e_{i}
\label{main_eq1}\\
w_{1i} & =x_{i}+u_{1i}
\label{main_eq2}\\
w_{ki} & =h_{k}(x_{i})+u_{ki},\ k=2,\ldots,K 
\label{main_eq3}
\end{align}
where the residual $e_i$ in \eqref{main_eq1} follows an unspecified distribution that satisfies \eqref{QR}. The measurement errors $u_{ki}$ in \eqref{main_eq2} and \eqref{main_eq3} are assumed to have a zero mean $E(u_{ki})=0$ and a constant variance $Var(u_{ki})=\sigma^2_k$, and are further assumed to be independent of each other and distributed independently of $x_{i}$ and $y_i|x_i$. Equation \eqref{main_eq2} implicitly assumes a reference variable that is only exposed to sampling or additive measurement errors. This assumption is natural and popular when survey datasets exist \citep{fuller2009sampling, berg2021} and can be generalized to other deterministic functions of proxy and unobserved covariates, such as logarithms \citep{berry2002bayesian}. We leave the other proxies $w_{k}$ $(k \ge 2)$ in \eqref{main_eq3} with an arbitrary function $x$, denoted by $h_k(x)$. 

\cite{lubotsky2006interpretation} used similar assumptions with \eqref{main_eq1} to \eqref{main_eq3} to estimate the regression coefficient for the linear regression of $y_i$ on $x_i$, with multiple proxies for $x_i$. They allowed nonzero covariance between the measurement errors; however, they assumed a linear relationship with all proxy variables $w_k$ for $k\ge 2$ in \eqref{main_eq3} and offered a lower bound on the regression coefficient of $y_i$ on $x_i$. Variable $w_1$ with additive error is the benchmark variable. The assumption of the existence of a benchmark variable is prevalent and necessary because it amounts to fixing the scale of unobserved $x$. 

Intuitively, the proxy variables are not required to be independent of each other or $x_i$. However, because measurement errors $u_{ki}$ are independent of each other and $x_i$ and $y_i|x_i$, the proxies are conditionally independent considering the true covariate $x_i$. This model setup makes the conditional distribution between observed proxies independent of each other; this suggests  that the likelihood of multiple proxy variables can be separately specified with a mixture representation based on conditional distribution given $x_i$ and their prior probability distribution. This makes Bayesian Gibbs sampling reasonable for the estimation method when $x_i$ is treated as an auxiliary variable and contributes to the generation of the observed data $\left(y_i,w_{1i},\ldots,w_{Ki}\right)$.

\section{Bayesian Estimation}\label{our_method}

This section briefly introduces the natural cubic spline and P-spline to be used for the quantile function $g_p(x)$ in \eqref{main_eq1}. The proposed methodology for combining multiple proxies in Bayesian quantile regression is followed by explicit sampling steps.

\subsection{Splines for Quantile Regression}\label{regression_splines}
Although the proposed framework is applicable to other quantile regression functions, such as the polynomial function of covariate $x$ \citep{koenker1978regression, yu2001bayesian, yu2003quantile}, polynomial quantile regression is frequently restricted because the degree of the polynomial must be chosen in advance, and data might have a limited local effect on the shape of the polynomial regression curve, especially for high quantiles \citep{thompson2010bayesian}. To alleviate the parametric assumption and secure the model’s flexibility, we consider two popular spline methods for $g_p(x)$: nonparametric cubic spline and the P-spline function. For details regarding the regression spline, refer to \cite{green1993nonparametric, eliers1996flexible, carroll1999nonparametric, brezger2006generalized}. 

\subsubsection{Natural Cubic Spline}\label{NCS}
The natural cubic spline is a piecewise cubic polynomial function with continuous first and second derivatives at each knot, and is linear beyond the boundary knots \citep{green1993nonparametric}. Suppose that $\tau_{1},\ldots,\tau_{N}$ are $N$ fixed knots covering a range of $x$ and $\boldsymbol{g}=\left(g_{1},\ldots,g_{N}\right)^{T}$ denote the values of the natural cubic spline $g_p(x)$ at knots $\tau_{1},\ldots\tau_{N}$, $g_p(\tau_{1}),\ldots g_p(\tau_{N})$. As a desirable property of the natural cubic spline, there is a unique natural cubic spline function $g_p(x)$ with knots $\tau_{1},\ldots,\tau_{N}$ satisfying $g(\tau_{i})=g_{i},\ i=1,\ldots,N$ for any given value $g_{1},\ldots,g_{N}$. Therefore, we can handle function $g(x)$ using its finite-length surrogate $\boldsymbol{g}$. In terms of Bayesian inference, we can model $g_p(x)$ by giving the prior to $\boldsymbol{g}$ and not $g_p(x)$. Following \cite{green1993nonparametric}, the prior for $\boldsymbol{g}$ is defined by a multivariate normal distribution as follows:
\begin{align*}
\pi\Big(\boldsymbol{g}|\lambda\Big) \propto \text{exp\ensuremath{\left(-\frac{1}{2}\lambda\boldsymbol{g}^{T}K\boldsymbol{g}\right)}}
\addtocounter{equation}{1}\tag{\theequation}, \label{g_prior}
\end{align*}	
where $K$ is the $N \times N$ matrix with rank$(K)=N-2$, a function of the difference between the knots defined by \citep{eubank1999nonparametric}. $\lambda$ contributes to the smoothness of curve $g$ and has a standard conjugate gamma prior:
\[
\pi\Big(\lambda\Big) = \frac{\lambda^{a_\lambda-1}\text{exp}(-\frac{\lambda}{b_\lambda})}{\Gamma(a_\lambda)b_\lambda^{a_\lambda}},\text{ }\lambda>0.
\]
The quadratic term $\boldsymbol{g}^{T}K\boldsymbol{g}$ in the exponential kernel is equivalent to the roughness penalty, $\int_{a}^{b}g_p^{''}(t)^{2}dt=\boldsymbol{g}^{T}K\boldsymbol{g}$ \citep{green1993nonparametric}. This type of quadratic prior, $\exp \left(-\frac{1}{2}\lambda\int g_p^{''}(t)^{2}dt\right)$, which measures the complexity of the parameter, is a natural choice because it corresponds to the penalty term in the penalized maximum likelihood. With this prior, the posterior log density of the function $g_p(x_i)$ is equal to the loss function in the regression context, with the roughness penalty added to the kernel of the log-likelihood function \citep{hoerl1970ridge, green1993nonparametric, tibshirani1996regression}.

The final step in the Bayesian approach is defining the natural cubic spline function’s likelihood by changing the conventional polynomial part of the standard quantile regression likelihood to the natural cubic spline form \citep{yu2001bayesian, thompson2010bayesian}. The resulting likelihood takes the following form:
\begin{align*}	
L\Big(y \mid \boldsymbol{g}, x\Big) = p^{n}(1-p)^{n}\text{exp} \left\{ -\sum_{i=1}^{n} \rho_{p}(y_{i}-g_p(x_{i}))\right\} 
\addtocounter{equation}{1}\tag{\theequation} \label{QR_likeli}
\end{align*} 
Notably, we explicitly specify $x$ in likelihood $L(y|\boldsymbol{g},x)$ for generalization, where $x$ is also an unknown variable. 

\subsubsection{Regression P-Spline}\label{BQR_section}

Another general approach to spline fitting is a penalized spline or simply a P-spline. For a P-spline of degree $l$ with $N$ fixed knots, $g_p(x)$ is defined by $\boldsymbol{Z}(x)^T\boldsymbol{\beta}$ where $\boldsymbol{Z}(x)$ is $(N+l+1)$ vector composed of B-spline basis functions evaluated at observation $x$, and $\boldsymbol{\beta}$ is the coefficient of the basis functions \citep{eliers1996flexible}. A conventional basis is $\boldsymbol{Z}(x)=\left(1,x,\ldots,x^l,(x-\tau_1)^l_+,\ldots,(x-\tau_N)^l_+\right)^T$. Then, $\beta_{2+l},\ldots,\beta_{1+N+l}$ are the sizes of the jumps in the $l$th derivative of $g(x)$ at the knots. 

\cite{eliers1996flexible} suggested a roughness penalty based on differences of adjacent spline coefficient to guarantee sufficient smoothness. In Bayesian analysis, the prior of $\boldsymbol\beta$ replaces the roughness penalty term of the penalized likelihood as their stochastic analogs \citep{lang2004bayesian}. Assuming a first-order random walk for $\boldsymbol \beta$, that is, 
\begin{align*}	
\pi\Big(\beta_k|\beta_{k-1},\lambda\Big) \propto N\Big(\beta_{k-1},\lambda\Big)
\end{align*} 
the joint conditional distribution of $\boldsymbol{\beta}$ is 
\begin{align*}	
\Big(\boldsymbol{\beta}|\lambda\Big)\propto \exp\left(-\frac{1}{2}\lambda\boldsymbol{\beta}^T K\boldsymbol{\beta}\right)
\end{align*} 
where $K=\lambda R^TR$ is $(N+l+1) \times (N+l+1)$ penalty matrix with rank$(K)=N+l$ and $R$ is a first-order difference matrix \citep{lang2004bayesian,waldmann2013bayesian}. This prior on $\boldsymbol{\beta}$ induces a prior on $g_p$ owing to the deterministic relationship between $g_p$ and $\boldsymbol{\beta}$, $g_p(x)=\boldsymbol{Z}(x)\boldsymbol{\beta}$. The precision parameter $\lambda$, again, contributes to the smoothness of curve $g$ and has a standard gamma prior:
\[
\pi\Big(\lambda\Big) = \frac{\lambda^{a_\lambda-1}\text{exp}(-\frac{\lambda}{b_\lambda})}{\Gamma(a_\lambda)b_\lambda^{a_\lambda}},\text{ }\lambda>0.
\]
The P-spline function’s likelihood can be defined by changing the conventional polynomial part of the standard quantile regression likelihood. The same result is obtained in \eqref{QR_likeli}. However, unlike the natural cubic spline, \cite{waldmann2013bayesian} suggested exploiting the stochastic representation of the likelihood for more efficient sampling \citep{kozumi2011gibbs}.

\subsection{Nonparametric Quantile Regression with Multiple Proxy Variables}

In this section, we describe the fully Bayesian approach to the problem setup in \eqref{main_eq1}–\eqref{main_eq3}. The unknown parameters to be estimated $\boldsymbol{\Theta} =\left(\boldsymbol{g},\lambda,\boldsymbol{x},\boldsymbol{\theta},\boldsymbol{\sigma^{2}}\right)$, where $\boldsymbol{\theta} =\left(\boldsymbol{\theta_{1},\ldots\theta_{K}}\right)^T,\ \boldsymbol{\sigma^{2}} =\left({\sigma_{1}^{2}, \ldots, \sigma_{K}^{2}}\right)^T$ and $\boldsymbol{\theta_k}$ is a parameter related to $h_k(x)$. Without the loss of generality, the posterior density is 
\begin{equation*}
p\left(\boldsymbol{\Theta}|y,\boldsymbol{w}\right) \propto p\left({y}|\boldsymbol{g},\lambda,{x}\right)p\left(\boldsymbol{w}|{x},\boldsymbol{\theta},\boldsymbol{\sigma^{2}}\right)
 \pi\left(\boldsymbol{g}|\lambda\right)\pi\left(\lambda\right)\pi\left(\boldsymbol{\theta}\right)\pi\left(\boldsymbol{\sigma^{2}}\right)\pi\left({x}\right)
\end{equation*}
where $\boldsymbol{w}=\left(w_1,\ldots,w_K\right)$. For the variance parameter, we use the conjugate independent inverse gamma prior $\pi(\boldsymbol{\sigma^{2}}) =\prod_{k=1}^K\pi(\sigma_{k}^{2})$. For the prior distribution of $\boldsymbol{x}$, the reasonable prior distribution and appropriate prior varies depending on the application \citep{berry2002bayesian}. For the exemplary prior, we use a widely used hierarchical normal distribution, that is, 
\begin{equation*}
\pi\left(x\right) \sim\mbox{N}(\mu_{x},\sigma_{x}^{2}),\quad
\mu_{x} \sim\mbox{N}(0,\sigma_{\mu}^{2}),\quad
\sigma_{x}^{2} \sim\mbox{IG}\Big(a_{x},b_{x}\Big)
\end{equation*}

From these prior settings, we derive the complete conditional distributions for $\boldsymbol x$ up to the normalizing constant 
\begin{align*}
\pi\Big(x_{i}\mid\Theta_{x-}\Big) & \propto\exp\left(-\rho_{p}\Big(y_{i}-g_{p}(x_{i})\Big)-\frac{1}{\sigma_{x}^{2}}\left(x_{i}-\mu_{x}\right)^{2}+\sum_{k=1}^{K}\left(-\frac{1}{\sigma_{k}^{2}}\Big(w_{ki}-h_{k}(x_{i})\Big)^2\right)\right)
\end{align*}
where $\Theta_{x-}$ denotes all other parameters except $x$. Other conditional distributions are derived from the conjugacy of their prior as follows: 
\begin{align*}
\pi\left(\sigma_{k}^{2}\mid\Theta_{\sigma_{k}-}\right) & \sim\text{IG}\left(\frac{n}{2}+a_{k},b_{k}+\frac{1}{2}\sum_{i=1}^{n}\Big(w_{ki}-h_{k}(x_{i})\Big)^{2}\right)\\
\pi\left(\sigma_{x}^{2}\mid\Theta_{\sigma_{x}-}\right) & \sim\text{IG}\left(\frac{n}{2}+a_{x},b_{x}+\frac{1}{2}\sum_{i=1}^{n}\left(x_{i}-\mu_{x}\right)^{2}\right)\\
\pi\left(\mu_{x}\mid\Theta_{\mu_{x}-}\right) & \sim\text{N}\left(\left(\frac{\sum_{i=1}^{n}x_{i}}{\sigma_{x}^{2}}\right)/\left(\frac{n}{\sigma_{x}^{2}}+\frac{1}{\sigma_{\mu}^{2}}\right),\left(\frac{n}{\sigma_{x}^{2}}+\frac{1}{\sigma_{\mu}^{2}}\right)^{-1}\right)
\end{align*}

For the parameters $\boldsymbol{\theta_k}$ related to the relationship $h_k(x)$ between $w_k$ and $x$, $h_k(x)$ can only contribute to its expectation of the conditional distribution, and the distribution’s family remains the same because $u_k$ is independent of $x$. Therefore, the generalization of $h_k(x)$ to an arbitrary functional form, such as a linear regression or natural cubic spline with a pre-existing Bayesian method for $\boldsymbol{\theta_k}$, is possible. We provide specific examples in the subsequent subsection.

The priors for spline-related parameters $\boldsymbol{g}$ and $\lambda$ differ depending on the regression spline function, as specified in \ref{regression_splines} and the detailed Gibbs sampling procedure for each spline function is specified in the subsequent subsection. 

A key  benefit of this Bayesian approach is that the smoothing spline’s observations are generated from the posterior; thus, we can estimate the entire posterior distribution of $g$, which was difficult in \cite{lubotsky2006interpretation}. 
Furthermore, an additional assumption is required to combine proxies and identify the unobserved covariate $x$ \citep{aigner1984latent, lubotsky2006interpretation}. However, in the Bayesian framework, treating unobserved $x$ as a latent variable and placing its prior probability distribution, which corresponds to the structural approach in the literature \citep{fuller2006measurement}, is natural.  
Although the regression function $g_p$ is the primary focus of interest, the joint posterior distribution is a powerful tool that enables the inference of the unobserved covariate $ x_i$ and its unobserved relationship with proxies $h_k(x)$.

\subsection{Gibbs Sampling Step}\label{sec:Gibbs}
The proposed framework for combining multiple proxies can be completed by defining the formula of $h_k$ and the distribution of the measurement error $u_{k}$, which can be defined by the researcher. We formulate the entire problem in the Bayesian framework using specific examples and present the Metropolis-Hastings steps \citep{gamerman2006markov}. 

\subsubsection{Natural Cubic Spline with Quadratic Proxy}

Suppose we have two proxies: one with an additive error and another with a quadratic relationship.
\begin{align*}
	y_{i} & =g_p(x_{i})+e_{i},\\
	w_{1i} & =x_{i}+u_{1i},\\
	w_{2i} & =\alpha_0+\alpha_1x_{i}+\alpha_2 x_i^2 + u_{2i},
	\addtocounter{equation}{1}\tag{\theequation} \label{ncs_example}
\end{align*}
where $u_{ki} \sim N(0,\sigma_{k}^{2}),\ k=1, 2$. Here, parameter $\boldsymbol{\theta_2}$ for $h_2(x)$ is $\boldsymbol{\alpha}=(\alpha_0,\alpha_1,\alpha_2)$. We use normal prior for $\pi(\alpha)\sim N(\mu_\alpha,V_\alpha)$. Following \cite{thompson2010bayesian}, we model a quantile function of covariate $x$ using the natural cubic spline, with $N$ evenly spaced fixed knots covering a range of $x$. 

A Gibbs sampling algorithm for the quantile regression model is constructed by sampling each component of $\Theta$ from the full conditional distributions. Following \cite{thompson2010bayesian}'s initialization, $\boldsymbol{g}^{(0)}$ is set as the posterior mean value of the quantile regression curve \citep{yu2001bayesian} at $\tau_{1},\dots\tau_{N}$, and $\lambda^{(0)}$ is obtained by applying generalized cross-validation of the usual smoothing spline \citep{green1993nonparametric}. Additionally, $x^{(0)}$ is set as a multiple proxies $w_{1}$ because it is a more reliable proxy in the initialization step with no information regarding $\boldsymbol{\alpha}$. 

One iteration of the Gibbs sampling algorithm at iteration $t$ is as follows:
\begin{enumerate}
\item Generate candidate $\boldsymbol{g}^{*}$ from the multivariate normal distributions, 
$$\boldsymbol{g}^{*} | \boldsymbol{g}^{(t-1)} \sim \text{MVN}(\boldsymbol{g}^{(t-1)},\Sigma),$$ 
and accept $\boldsymbol{g}^{*}$ with probability,
\[
r=\text{min}\left\{ 1,\frac{L(y|\boldsymbol{g}^{*},x^{(t-1)})\pi(\boldsymbol{g}^{*}|\lambda)q(\boldsymbol{g}^{(t-1)}|\boldsymbol{g}^{*})}{L(y|\boldsymbol{g}^{(t-1)},x^{(t-1)})\pi(\boldsymbol{g}^{(t-1)}|\lambda)q(\boldsymbol{g}^{*}|\boldsymbol{g}^{(t-1)})}\right\},
\]
where $q$ represents the proposal density function.
 
\item Generate candidate $\lambda^{*}$ from the log-normal distribution, $$\eta^{*}\sim\text{N}(\text{log}(\lambda^{(t-1)}),\sigma_{\lambda}^{2}),$$
where $\lambda^{*}=\exp(\eta^{*})$, and accept $\lambda^{*}$ with probability, 
\[
  r=\min\left\{ 1,\frac{\pi(\boldsymbol{g}^{(t)}|\lambda^{*})\pi(\lambda^{*})q(\lambda^{(t-1)}|\lambda^{*})}{\pi(\boldsymbol{g}^{(t)}|\lambda^{(t-1)})\pi(\lambda^{(t-1)})q(\lambda^{*}|\lambda^{(t-1)})}\right\}. 
\]

\item Generate $x^{*}$ from the multivariate normal distribution, $$x^{*}|x^{(t-1)}\sim\text{MVN}(x^{(t-1)},\Sigma_{xx}),$$
and accept $x^{*}$ with probability,
\[
 r =\min\left\{1, \frac{L(y|\boldsymbol{g}^{(t)},x^{*})\pi(w_{1}|w_{2},x^{*})\pi(w_{2}|x^{*})\pi(x^{*}|\mu_{x}^{(t-1)},\left(\sigma_{x}^{2}\right)^{(t-1)})q(x^{(t-1)}|x^{*})}{L(y|\boldsymbol{g}^{(t)},x^{(t-1)})\pi(w_{1}|w_{2},x^{(t-1)})\pi(w_{2}|x^{(t-1)})\pi(x^{(t-1)}|\mu_{x}^{(t-1)},\left(\sigma_{x}^{2}\right)^{(t-1)})q(x^{*}|x^{(t-1)})}\right\}. 
\]
From the independent assumption, this step can be separated as generating $x_{i}^{*}$ from the normal $x_{i}^{*}|x_{i}^{(t-1)}\sim N(x_{i}^{(t-1)},\sigma_{xx}^{2})$ with acceptance probability reduced to the term of the $i$-th data.
  
\item Sample $\left(\sigma_{1}^{2}\right)^{(t)}\sim\text{Inv-Gamma}\left(\frac{n}{2}+a_{1},b_{1}+\frac{1}{2}\sum_{i=1}^{n}\left(w_{1i}-x_{i}^{(t)}\right)^{2}\right)$, where $\text{Inv-Gamma}(a,b)$ indicates the inverse gamma with shape parameter $a$ and scale parameter $b$, and $a_{1}$ and $b_{1}$ are the corresponding parameters for the prior of $\sigma_{1}^{2}$.

\item Sample $\left(\sigma_{2}^{2}\right)^{(t)}\sim\text{Inv-Gamma}\left(\frac{n}{2}+a_{2},b_{2}+\frac{1}{2}\sum_{i=1}^{n}\left(w_{2i}-\left(\alpha_{0}^{(t-1)}+\alpha_{1}^{(t-1)}x_{i}^{(t)}\right)\right)^{2}\right)$ with $a_{2}$ and $b_{2}$ be the parameters for the prior of $\sigma_{2}^{2}$.

\item Sample $\left(\sigma_{x}^{2}\right)^{(t)}\sim\text{Inv-Gamma}\left(\frac{n}{2}+a_{x},b_{x}+\frac{1}{2}\sum_{i=1}^{n}\left(x_{i}^{(t)}-\mu_{x}^{(t-1)}\right)^{2}\right)$ with $a_{x}$ and $b_{x}$ be parameters for the prior of $\sigma_{x}^{2}$.

\item Sample $\mu_{x}^{(t)}\sim\text{N}(M_{*},V_{*})$, where 
\[
  V_{*}=\left(\frac{n}{\left(\sigma_{x}^{2}\right)^{(t)}}+\frac{1}{\sigma_{\mu}^{2}}\right)^{-1}\text{and }M_{*}=\left(\frac{\sum_{i=1}^{n}x_{i}^{(t)}+\frac{M_{\mu}}{\sigma_{\mu}^{2}}}{\left(\sigma_{x}^{2}\right)^{(t)}}\right)/\left(\frac{n}{\left(\sigma_{x}^{2}\right)^{(t)}}+\frac{1}{\sigma_{\mu}^{2}}\right)
\]
with $M_{\mu}$ and $\sigma_{\mu}^2$ the prior mean and variance for $\mu_{x}$.

\item Sample $\alpha^{(t)}\sim\text{MVN}(\mu_{*},V_{*})$, where 
\[
  V_{*}=\left(\frac{X^{T}X}{\left(\sigma_{2}^{2}\right)^{(t)}}+V_{\alpha}^{-1}\right)^{-1}\text{and }\mu_{*}=\left(\frac{X^{T}X}{\left(\sigma_{2}^{2}\right)^{(t)}}+V_{\alpha}^{-1}\right)^{-1}\left(\frac{X^{T}W_{2}}{\left(\sigma_{2}^{2}\right)^{(t)}}+V_{\alpha}^{-1}\mu_{\alpha}\right)
\]
with $M_{\alpha}$ and $V_{\alpha}$ as the prior mean vector and covariance matrix for $\alpha$ and $X$ as the vector of $x_{i}^{(t)}$s, $i=1,\dots,n$.
\end{enumerate}

Steps 1--3 require the Metropolis-Hastings algorithm, and the other steps can be sampled from the conjugate distribution. The inference regarding the unobserved regressor, quantile spline function, or regression coefficient is based on these posterior samples. 

\subsubsection{P-Spline with arbitrary nonlinear Proxy}
For most observed data in applications, prespecifying the relationship $h_k$ between the observed proxy and unobserved covariates in advance is challenging . Usually, a polynomial parametric relationship is applied with a certain purpose, such as interpretability. However, one might want  a more flexible model with few prior assumptions regarding the relationship between the proxy and unobserved covariates. Provided that a benchmark variable is available for fixing the scale of unobserved $x$, the proposed method can adopt a nonlinear relationship for the other $h_k$. 

Suppose we have two proxies---one with an additive error and another with a nonlinear relationship.
\begin{align*}
	y_{i} & =g_p(x_{i})+e_{i},\\
	w_{1i} & =x_{i}+u_{1i},\\
	w_{2i} & =h_2(x_i) + u_{2i},
	\addtocounter{equation}{1}\tag{\theequation} \label{bqr_example}
\end{align*}
where $u_{ki} \sim N(0,\sigma_{k}^{2}),\ k=1,2$. 

Following \cite{waldmann2013bayesian}, we used a stochastic representation of the likelihood $L\Big(y \mid \boldsymbol{g}, x\Big)$, that is, $y \mid \boldsymbol{g}, x \propto \mbox{N}(g_{p}(x)+As,B\frac{s}{\delta^{2}})$, where $A=\frac{1-2p}{p(1-p)},\ B=\frac{2}{p(1-p)}$ and $s \sim \text{Exp}(\delta^{2})$ with the conjugate gamma prior on $\delta^2\sim \mbox{GA}(a_\delta,b_\delta)$. This hierarchical representation of the likelihood in \eqref{QR_likeli} enables efficient Gibbs sampling. For additional details, refer to \cite{kozumi2011gibbs, waldmann2013bayesian}. 

As we have two nonlinear functions $g_p$ and $h_2$ to be fitted, we use two P-splines. To discriminate the coefficients of the basis function for $g_p$ and $h_2$, we use subscripts $\boldsymbol{\beta}_g$ and $\boldsymbol{\beta}_h$. For a P-spline with $N$ fixed knots, $g_p$ is defined by $\boldsymbol{\beta}_g$, and for any realization of $\boldsymbol{\beta}_g$, there exists a corresponding realization $g_p(x)=\boldsymbol {Z}(x)\boldsymbol{\beta}_g$. As both splines $g_p$ and $h_2$ share covariate $x$, they share the same knots and penalty matrix $K$, which reduces the computation. Thereafter, the prior is as specified in Section \ref{BQR_section}; $\pi(\boldsymbol{\beta}_g)\propto\exp \Big(-\frac{1}{2} \lambda_g \boldsymbol{\beta}_g^T K \boldsymbol{\beta}_g\Big)$, $\pi(\boldsymbol{\beta}_h)\propto\exp \Big(-\frac{1}{2} \lambda_h \boldsymbol{\beta}_h^T K \boldsymbol{\beta}_h\Big)$, $\pi(\lambda_g) \sim \mbox{GA}(a_{\lambda_g},b_{\lambda_g})$ and $\pi(\lambda_h) \sim \mbox{GA}(a_{\lambda_h},b_{\lambda_h})$.

One iteration of the Gibbs sampling algorithm at iteration $t$ is as follows:
\begin{enumerate}
\item Generate $x_{i}^{*}$ from the normal distribution, $x_{i}^{*}|x_{i}^{(t-1)}\sim\text{N}(x^{(t-1)},\sigma_{xx}^{2}),$
and accept $x_{i}^{*}$ with probability,
\begin{align*}
r & =\min\Big\{1,\frac{N(y_{i}|Z(x_{i}^{*})^{T}\beta_{g}^{(t-1)}+As_{i}^{(t-1)},B\frac{s_{i}^{(t-1)}}{\delta^{2^{(t-1)}}})}{N(y_{i}|Z(x_{i}^{(t-1)})^{T}\beta_{g}^{(t-1)}+As_{i}^{(t-1)},B\frac{s_{i}^{(t-1)}}{\delta^{2^{(t-1)}}})}\\
 & \times\frac{\pi(w_{1i}|x_i^{*})\pi(w_{2i}|\beta_{h}^{(t-1)},x_i^{*})\pi(x_i^{*}|\mu_{x}^{(t-1)},\left(\sigma_{x}^{2}\right)^{(t-1)})q(x_i^{(t-1)}|x_i^{*})}{\pi(w_{1i}|x_i^{(t-1)})\pi(w_{2i}|\beta_{h}^{(t-1)},x_i^{(t-1)})\pi(x_i^{(t-1)}|\mu_{x}^{(t-1)},\left(\sigma_{x}^{2}\right)^{(t-1)})q(x_i^{*}|x_i^{(t-1)})}\Big\}
\end{align*}

\item Sample $\boldsymbol{\beta_{g}}^{(t)}\sim\text{N}(M_{*},V_{*})$, where
\[
V_{*}=\left(\lambda_{g}^{(t-1)}K+\frac{\delta^{2^{(t-1)}}}{B}Z^{T}D^{-1}Z\right)^{-1}\text{and }M_{*}=V_{*}^{-1}\left(\frac{\delta^{2^{(t-1)}}}{B}Z^{T}D^{-1}\left({y}-A{s}^{(t-1)}\right)\right)
\]
with $D=\text{diag}(s_{1}^{(t-1)},\ldots s_{n}^{(t-1)})$ and $Z$ represents the design matrix with $Z(x_{i}^{(t)}),\ i=1,\ldots,n$.

\item Sample $\lambda_{g}^{(t)}\sim\text{GA}(a_{\lambda}+0.5\text{rank}(\lambda_{g}^{(t-1)}K),b_{\lambda}+0.5\boldsymbol{\beta_{g}}^{(t)T}\lambda_{g}^{(t-1)}K\boldsymbol{\beta_{g}}^{(t)})$, where $\text{GA}(a,b)$ is gamma distribution with shape parameter $a$ and rate parameter $b$, and $a_{\lambda_{g}}$ and $b_{\lambda_{g}}$ are the corresponding parameters for the prior of $\lambda_{g}$.

\item Sample $s_{i}^{{-1}^{(t)}}\sim\text{Inv-Gauss}\left(\frac{\sqrt{A^{2}+2B}}{y_{i}-Z(x_{i}^{(t)})^{T}\boldsymbol{\beta_{g}^{(t)}}},\frac{\delta^{2^{(t-1)}}(A^{2}+2B)}{B}\right)$, where $\text{Inv-Gauss}(a,b)$ is an inverse gaussian distribution with mean parameter $a$ and shape parameter $b$. 

\item Sample $\delta^{2^{(t)}}\sim GA(a_{\delta}+\frac{3n}{2},b_{\delta}+\frac{1}{2B}\sum_{i=1}^{n}s_{i}^{(t)^{-1}}\left(y_{i}-Z(x_{i}^{(t)})^{T}\beta_{g}^{(t)}-As_{i}^{(t)}\right)^{2}+\sum_{i=1}^{n}s_{i}^{(t)})$, where $a_{\delta}$ and $b_{\delta}$ are the corresponding parameters for the prior of $\delta^{2}$.

\item Sample $\mu_{x}^{(t)}\sim\text{N}(M_{*},V_{*})$, where 
\[
V_{*}=\left(\frac{n}{\left(\sigma_{x}^{2}\right)^{(t-1)}}+\frac{1}{\sigma_{\mu}^{2}}\right)^{-1}\text{and }M_{*}=\left(\frac{\sum_{i=1}^{n}x_{i}^{(t)}+\frac{M_{\mu}}{\sigma_{\mu}^{2}}}{\left(\sigma_{x}^{2}\right)^{(t-1)}}\right)/\left(\frac{n}{\left(\sigma_{x}^{2}\right)^{(t-1)}}+\frac{1}{\sigma_{\mu}^{2}}\right)
\]
with $M_{\mu}$ and $\sigma_{\mu}^{2}$ the prior mean and variance for $\mu_{x}$.

\item Sample $\left(\sigma_{x}^{2}\right)^{(t)}\sim\text{Inv-Gamma}\left(\frac{n}{2}+a_{x},b_{x}+\frac{1}{2}\sum_{i=1}^{n}\left(x_{i}^{(t)}-\mu_{x}^{(t)}\right)^{2}\right)$ with $a_{x}$ and $b_{x}$ as the parameters for the prior of $\sigma_{x}^{2}$

\item Sample $\left(\sigma_{1}^{2}\right)^{(t)}\sim\text{Inv-Gamma}\left(\frac{n}{2}+a_{1},b_{1}+\frac{1}{2}\sum_{i=1}^{n}\left(w_{1i}-x_{i}^{(t)}\right)^{2}\right)$ with $a_{1}$ and $b_{1}$ as the parameters for the prior of $\sigma_{1}^{2}$.

\item Sample $\boldsymbol{\beta_{h}}^{(t)}\sim\text{N}(M_{*},V_{*})$, where
\[
V_{*}=\left(\lambda_{h}^{(t-1)}K+\frac{1}{\sigma_{2}^{2}}Z^{T}Z\right)^{-1}\text{and }M_{*}=V_{*}^{-1}\left(\frac{1}{\sigma_{2}^{2}}Z^{T}{w_{1}}\right)
\]

\item Sample $\lambda_{h}^{(t)}\sim\text{GA}(a_{\lambda_{h}}+0.5\text{rank}(\lambda_{h}^{(t-1)}K),b_{\lambda_{h}}+0.5\boldsymbol{\beta_{h}}^{(t)T}\lambda_{h}^{(t-1)}K\boldsymbol{\beta_{h}}^{(t)})$, where $a_{\lambda_{h}}$ and $b_{\lambda_{h}}$ are the corresponding parameters for the prior of $\lambda_{h}$. 

\item Sample $\left(\sigma_{2}^{2}\right)^{(t)}\sim\text{Inv-Gamma}\left(\frac{n}{2}+a_{2},b_{2}+\frac{1}{2}\sum_{i=1}^{n}\left(w_{2i}-Z(x_{i}^{(t)})^{T}\beta_{h}^{(t)}\right)^{2}\right)$ with $a_{2}$ and $b_{2}$ as the parameters for the prior of $\sigma_{2}^{2}$.
\end{enumerate}

The inference regarding the unobserved regressor, quantile spline function, or regression coefficient is based on these posterior samples.

\section{Simulation}\label{simulation}
We conduct a simulation study to empirically evaluate the proposed method using various datasets. This simulation has the following three purposes: to evaluate the flexibility of the proposed method in various datasets by considering several different error types, to compare the proposed method with an alternative approach, and to evaluate the effect of different types of proxies and effect of the number of proxies on the proposed method.

\begin{figure}[htbp]
\centering	
\begin{tabular}{ccc}
(a) & (b) & (c)\\
\includegraphics[scale=0.6]{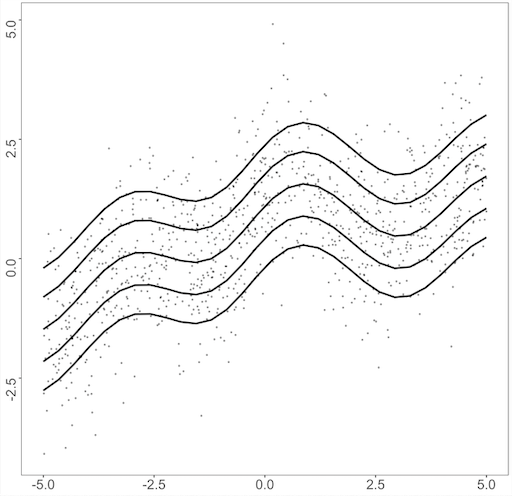} &
\includegraphics[scale=0.6]{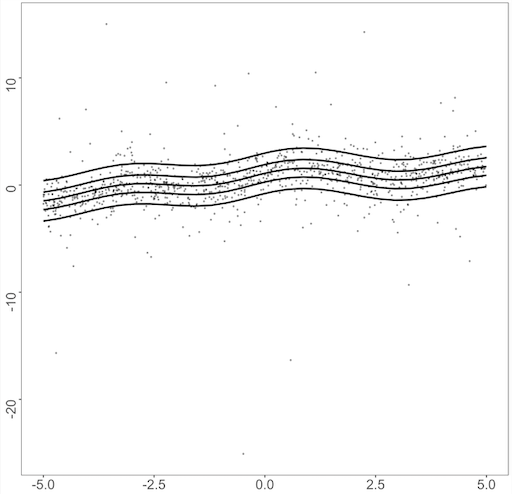} &
\includegraphics[scale=0.6]{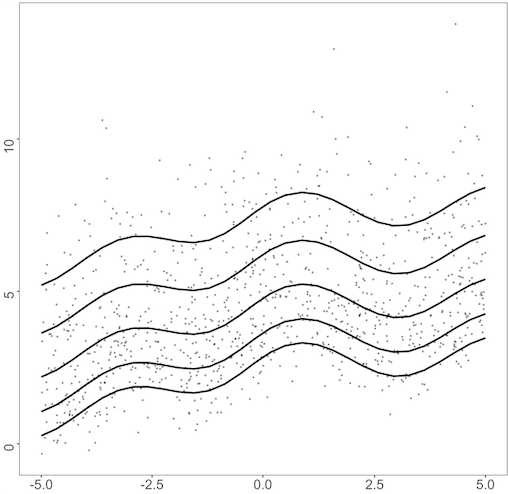} \\
(d) & (e) & (f)\\
\includegraphics[scale=0.6]{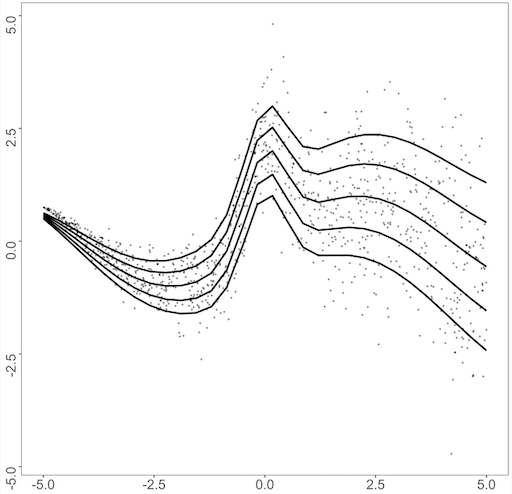} &
\includegraphics[scale=0.6]{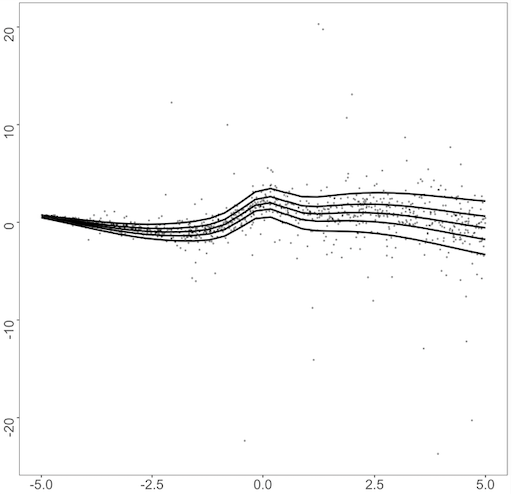} &
\includegraphics[scale=0.6]{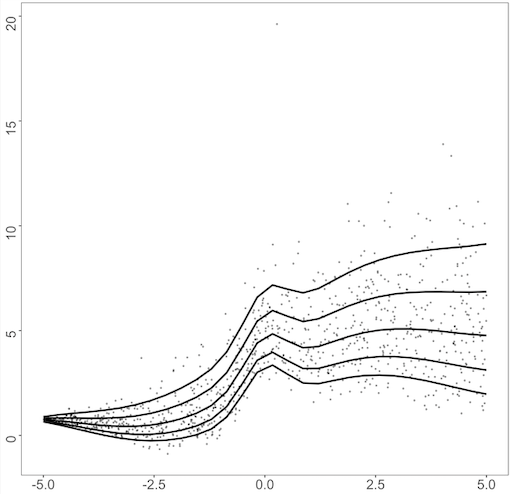}  \\
\end{tabular}
\caption{Simulated examples of Dataset1 (top) and Dataset2 (bottom) with standard normal distribution (left), Student $t$ distribution (middle), and gamma distribution (right). The black lines represent true quantile line for $p\in\{0.1,\ 0.25,\ 0.5,\ 0.75,\ 0.9\}$, respectively. }
\label{fig:exmaple_figure}
\end{figure}

We use the dataset studied by \cite{yue2011bayesian,waldmann2013bayesian}. To match the scale between each dataset, we scale the range of $x$ to $\left[-5,5\right]$ for each simulation using the appropriate affine transformation. We simulate the datasets using the following formulae:

\begin{itemize}
\item Dataset1 : $y_{i}=0.4x_{i}+0.5\sin(2.7x_{i})+1.1/(1+x_{i}^{2})+e_{i}$;
\item Dataset2 : $y_{i}=\sin(2(4x_{i}-2))+2\exp((-16^{2})(x-0.5)^{2})+((3x_{i})/2)e_{i}$,
\end{itemize}
and the quantile functions for each dataset are given by 
\begin{itemize}
\item Dataset1 : $g_{p}(x_{i})=0.4x_{i}+0.5\sin(2.7x_{i})+1.1/(1+x_{i}^{2})+F^{-1}(p)$
\item Dataset2 : $g_{p}(x_{i})=\sin(2(4x_{i}-2))+2\exp((-16^{2})(x-0.5)^{2})+((3x_{i})/2)F^{-1}(p)$
\end{itemize}
where $F^{-1}$ is the cumulative distribution function of the distribution from where the error $e_{i}$ is sampled.
For the error distribution $F$, we consider three different error distributions as follows: standard normal distribution, Student’s $t$ distribution with two degrees of freedom, and gamma distribution with shape 4 and rate 1. Figure \ref{fig:exmaple_figure} displays a data structure of both datasets with each error term, on $p\in\{0.1,\ 0.25,\ 0.5,\ 0.75,\ 0.9\}$. Dataset2 follows the heteroscedastic structure, as the error generated from the error distribution is multiplied by $x_{i}$, and the resulting quantile curves are no longer parallel to each other as in Dataset1. For the error distribution, $t$ is a heavy-tailed distribution, which may cause the dataset to include extreme outliers. The gamma distribution with shape 4 and rate 1 has a nonzero expectation skewed to the right. This causes the resulting quantile function to shift to positive values for higher quantiles. Similar examples were analyzed in \cite{kottas2009bayesian, taddy2010bayesian, fenske2011identifying, yue2011bayesian, waldmann2013bayesian}.

The proposed method can adapt multiple proxies with arbitrary relationships to the covariates. To assess a proxy’s effect, further, we suppose that the actual covariates $x$ are not directly observed, and that proxies $w_{k},\ k=1,\ 2,\ 3$ are observed. For the relationship $h_{k}$ between proxy $w_{k}$ and unobserved covariate $x$, we consider three different types as follows: identity, polynomial, and smooth nonlinear functions. That is, we have 
\begin{align*}
w_{1i} & =x_{i}+u_{1i}\\
w_{2i} & =\alpha_{0}+\alpha_{1}x_{i}+\alpha_{2}x_{i}^{2}+u_{2i}\\
w_{3i} & =\sin(12(x_{i}+0.1))/(x_{i}+0.1)+u_{3i}
\end{align*}
with $\boldsymbol{\alpha}=(3,\ 0.25,\ 0.75)$ and $u_{ki}\sim N(0,1),\ k=1,2,3$. The nonlinear example $h_3$ is based on \cite{friedman2001elements} and is generated from smoothing splines. The parameter for the quadratic coefficient $\boldsymbol{\alpha}$ is determined such that the ratio of variance in the error components to the total variance in the proxy variables in $w_2$ is roughly matched to that in $w_3$.
 
Most studies that have considered the regression problem in the presence of a proxy variable with an additive error assumed the existence of replicates of the benchmark variable, which is a stricter assumption than the case considered here \citep{carroll2006measurement, wei2009quantile}. Most studies have focused on mean regression; few have studies have considered estimating the quantile function in the case of proxies. Instead, we adjust \cite{carroll1999nonparametric}'s approach to be suitable for quantile regression problems. The method uses a partially Bayesian approach, which estimates the moment function of unknown $x$ using $w$ and estimates the spline function by minimizing the conventional penalized likelihood with the given estimated moments. This method uses a two-step estimation procedure that uses the information of $w$ to estimate $x$, and the relationship between $x$ and $y$ is estimated thereafter. 

\begin{figure}[htbp]
\centering	
\begin{tabular}{ccc}
(a) & (b) \\
\includegraphics[width=0.5\textwidth]{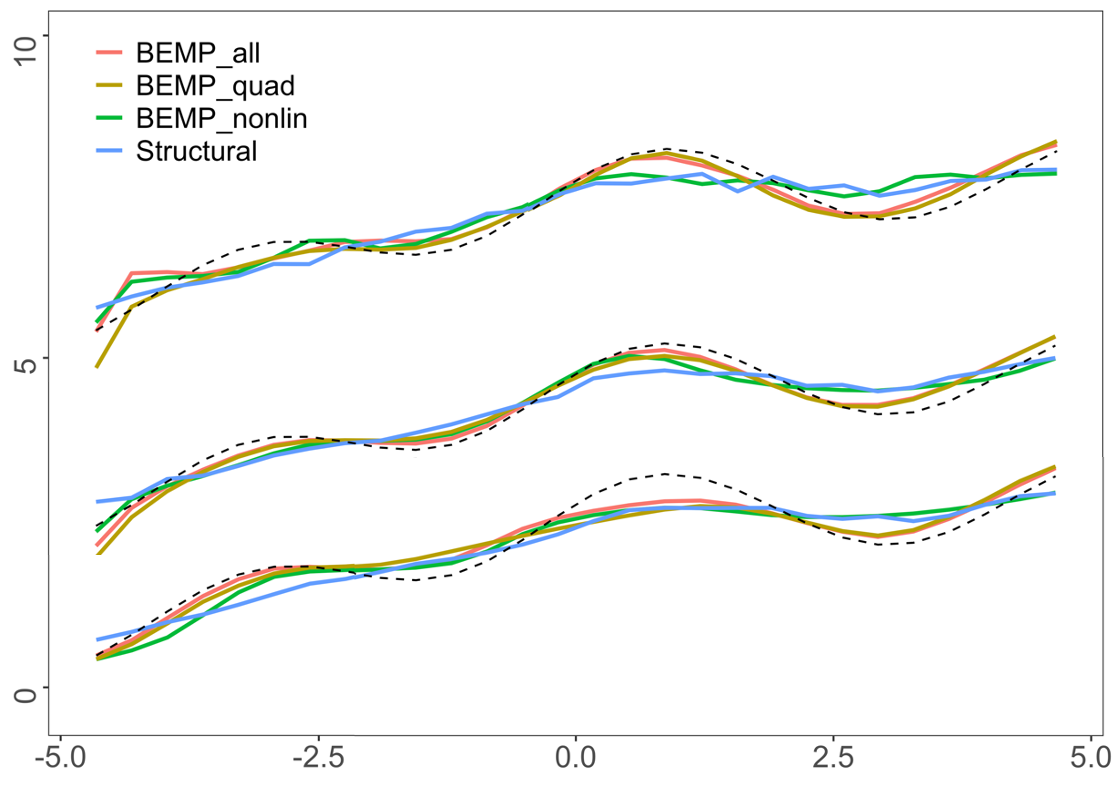} &
\includegraphics[width=0.5\textwidth]{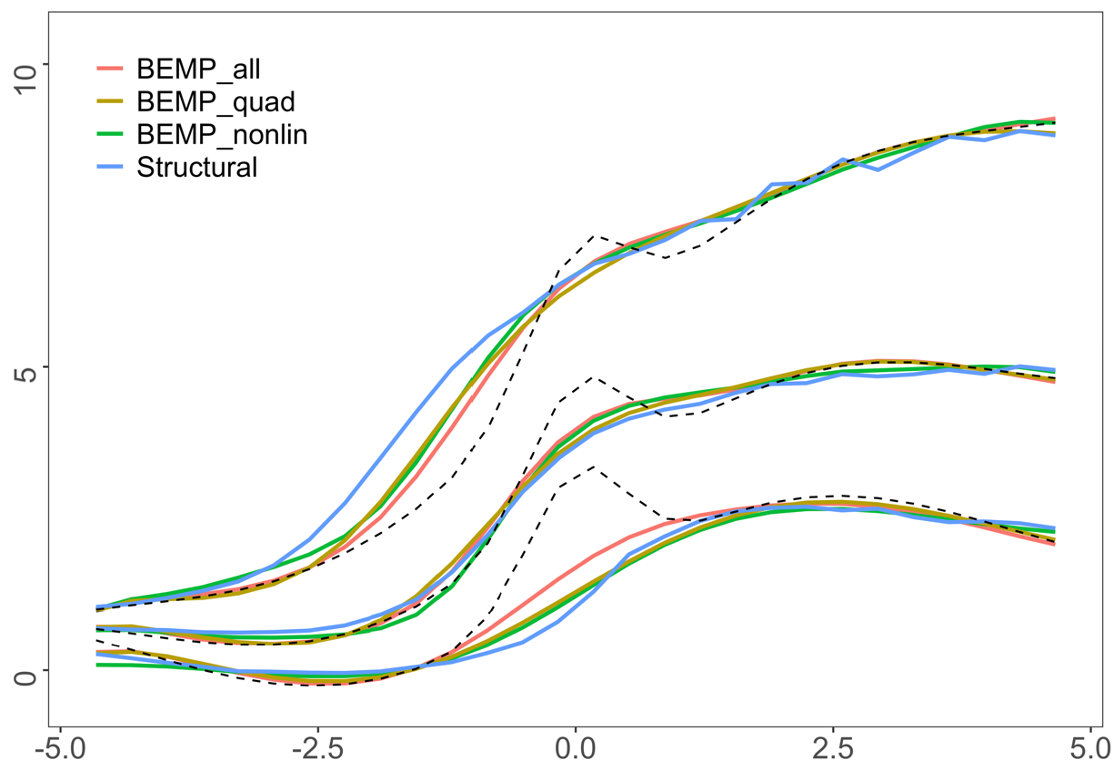} \\
\end{tabular}
\caption{Fitted line for estimators in Dataset1 (a) and Dataset2 (b), with gamma distributed error in $p_0=0.1,\ 0.5,\ 0.9$. The dotted black line represents the true quantile function. 
}
\label{fig:fitted_res}
\end{figure}

Consequently, five estimators were considered for comparison in each simulation.
\begin{enumerate}
\item Model without measurement error (woME): the benchmark estimator with true covariate $x$ directly observed without error. 
\item Structural estimator (Structural): the estimator calculated using \cite{carroll1999nonparametric}'s model, with all related prior settings equal to Bayesian estimator with multiple proxies (BEMP).
\item Bayesian estimator with polynomial proxies (BEMP-poly): the proposed Bayesian estimator that incorporates two proxies $w_{1}$ and $w_{2}$, but not $w_{3}$.
\item Bayesian estimator with nonlinear proxies (BEMP-nonlinear): the proposed Bayesian estimator that incorporates two proxies $w_{1}$ and $w_{3}$, but not $w_{2}$.
\item Bayesian estimator with all proxies (BEMP-all): the proposed Bayesian estimator that incorporates all proxies $w_{1}$, $w_{2}$, and $w_{3}$.
\end{enumerate}
For the woME model, we used the natural cubic spline method applied in quantile regression using \cite{thompson2010bayesian}'s model. For conciseness and unity, we present only the performance of the BEMP with a natural cubic spline. The performance of a BEMP with a P-spline is similar to that of a BEMP with a natural cubic spline, and is presented in the Appendix. The computation of the generation of $\boldsymbol g$ depends on the size of knots $N$. However, \cite{eliers1996flexible} and \cite{ruppert2002selecting} reported that computation can be reduced by using fewer knots $N\ll n$ with no loss of precision, and generally, $N$ between 20 and 40 ensures sufficient flexibility. We used the same 30 knots as those in \cite{thompson2010bayesian}.

For each of the generated datasets, the number of observations is fixed at $n=1000$ and the quantile functions on a fixed grid $p\in\{0.1,\ 0.25,\ 0.5,\ 0.75,\ 0.9\}$ are estimated. We assume identical MCMC sampling settings for all the five models. We set the number of iterations to 300,000 and took every 50th sample after discarding the first 50,000 steps as the burn-in period. The convergence is satisfactory, and an average of 5,000
posterior samples are used for the point estimation. After running 100 Monte Carlo (MC) simulations, we report their average and standard deviation.

We use two popular metrics to compare the estimators \citep{hardle1986approximations, fan1992design, gelfand1998model}:
\begin{itemize}
\item Mean squared error (MSE): 
\[
MSE(\hat{g}_{p})=\frac{1}{n}\sum_{i=1}^{n}\left(g_{p}(x_{i})-\hat{g}_{p}(x_{i})\right)^{2}
\]
where $\hat{g}_{p}(x_{i})$ is the estimate of ${g}_{p}(x_{i})$.
\item Posterior predictive loss (PPL):
\[
PPL_{m}(\hat{g}_{p})=\sum_{i=1}^{n}\sigma^2_{\hat{g}_{p}}(x_i)+\frac{m}{m+1}\sum_{i=1}^{n}\left(g_{p}(x_{i})-\hat{g}_{p}(x_{i})\right)^{2}
\]
where $\sigma^2_{\hat{g}_{p}}(x_i)$ is the posterior predictive distribution’s variance for ${g}_{p}(x_{i})$.
\end{itemize}
The first component in the PPL is a penalty term for model complexity, and the second is a term for goodness-of-fit. For the weight term $m$, we use $m=1,\ \infty$.

\subsection{Simulation Result}\label{simulation_result}
Table \ref{tab:dataset1_normal} summarizes the simulation results for Dataset1 with normal errors: All the models that we test exhibit superior performance to the naive model, which directly treats the proxy with error $w_2$ as a true covariate (see Appendix). The proposed BEMP outperforms the comparative method across all metrics. For the models using two proxies, BEMP-poly outperforms BEMP-nonlinear, though we matched the ratio of variance in the error components to the total variance in the proxy variables. This is a predictable result because BEMP-nonlinear uses two sets of spline parameters, and the number of parameters to be estimated may adversely affect the resulting performance  \citep{friedman2001elements}. However, notably, BEMP-all exhibits the most optimal performance, even better than BEMP-poly. The superior performance of BEMP-all is interesting because the number of parameters in BEMP-all is larger than that in BEMP-nonlinear. We believe that this is because when the amount of proxy used increases, the increased amount of information aids the estimation of each other  and provides a more precise estimation. In other words, a more precise estimation of $x$ makes it easier to estimate the spline parameter and vice versa. A similar discussion can be found in \cite{berry2002bayesian, lubotsky2006interpretation}, which argued for the effect of using multiple types of information. We conduct additional simulations to validate this assumption.

\begin{table}[htbp]
\centering
\caption{Monte Carlo means (standard errors) of MSE, $\text{PPL}_{1}$, and $\text{PPL}_{\infty}$ for homogeneous Dataset1 with standard normal distributed error}
\label{tab:dataset1_normal}
\begin{tabular}{c|c|ccccc}
\hline 
\multicolumn{1}{c}{{\small{}Quantile}} & \multicolumn{1}{c}{} & {\small{}woME} & {\small{}Structural} & {\small{}BEMP-poly} & {\small{}BEMP-nonlinear} & {\small{}BEMP-all}\tabularnewline
\hline 
\multirow{3}{*}{{\small{}$p=0.1$}}
  & {\small{}MSE} & {\small{}0.029 (0.013)} & {\small{}0.163 (0.038)} & {\small{}0.09 (0.031)} & {\small{}0.146 (0.06)} & \textbf{\small{}0.063 (0.027)}\tabularnewline
 & {\small{}$\text{PPL}_{1}$} & {\small{}0.08 (0.009)} & {\small{}0.204 (0.041)} & {\small{}0.115 (0.019)} & {\small{}0.137 (0.034)} & \textbf{\small{}0.1 (0.017)}\tabularnewline
 & {\small{}$\text{PPL}_{\infty}$} & {\small{}0.094 (0.015)} & {\small{}0.287 (0.053)} & {\small{}0.161 (0.033)} & {\small{}0.214 (0.062)} & \textbf{\small{}0.132 (0.028)}\tabularnewline
\hline 
\multirow{3}{*}{{\small{}$p=0.25$}}
  & {\small{}MSE} & {\small{}0.019 (0.008)} & {\small{}0.129 (0.03)} & {\small{}0.053 (0.018)} & {\small{}0.097 (0.035)} & \textbf{\small{}0.035 (0.015)}\tabularnewline
 & {\small{}$\text{PPL}_{1}$} & {\small{}0.05 (0.005)} & {\small{}0.15 (0.024)} & {\small{}0.069 (0.009)} & {\small{}0.089 (0.019)} & \textbf{\small{}0.061 (0.009)}\tabularnewline
 & {\small{}$\text{PPL}_{\infty}$} & {\small{}0.059 (0.009)} & {\small{}0.216 (0.034)} & {\small{}0.097 (0.017)} & {\small{}0.134 (0.035)} & \textbf{\small{}0.079 (0.016)}\tabularnewline
\hline 
\multirow{3}{*}{{\small{}$p=0.5$}}
  & {\small{}MSE} & {\small{}0.016 (0.007)} & {\small{}0.113 (0.029)} & {\small{}0.04 (0.017)} & {\small{}0.087 (0.04)} & \textbf{\small{}0.03 (0.014)}\tabularnewline
 & {\small{}$\text{PPL}_{1}$} & {\small{}0.038 (0.005)} & {\small{}0.124 (0.021)} & {\small{}0.053 (0.008)} & {\small{}0.071 (0.015)} & \textbf{\small{}0.045 (0.007)}\tabularnewline
 & {\small{}$\text{PPL}_{\infty}$} & {\small{}0.046 (0.008)} & {\small{}0.186 (0.032)} & {\small{}0.074 (0.016)} & {\small{}0.113 (0.034)} & \textbf{\small{}0.06 (0.013)}\tabularnewline
\hline 
\multirow{3}{*}{{\small{}$p=0.75$}}
  & {\small{}MSE} & {\small{}0.016 (0.01)} & {\small{}0.126 (0.032)} & {\small{}0.071 (0.027)} & {\small{}0.155 (0.035)} & \textbf{\small{}0.058 (0.038)}\tabularnewline
 & {\small{}$\text{PPL}_{1}$} & {\small{}0.04 (0.005)} & {\small{}0.143 (0.019)} & {\small{}0.063 (0.011)} & {\small{}0.088 (0.015)} & \textbf{\small{}0.055 (0.015)}\tabularnewline
 & {\small{}$\text{PPL}_{\infty}$} & {\small{}0.049 (0.009)} & {\small{}0.209 (0.03)} & {\small{}0.096 (0.025)} & {\small{}0.165 (0.032)} & \textbf{\small{}0.083 (0.033)}\tabularnewline
\hline 
\multirow{3}{*}{{\small{}$p=0.9$}}
  & {\small{}MSE} & {\small{}0.03 (0.026)} & {\small{}0.167 (0.044)} & {\small{}0.129 (0.025)} & {\small{}0.159 (0.034)} & \textbf{\small{}0.128 (0.036)}\tabularnewline
 & {\small{}$\text{PPL}_{1}$} & {\small{}0.059 (0.012)} & {\small{}0.203 (0.028)} & {\small{}0.084 (0.011)} & {\small{}0.094 (0.016)} & \textbf{\small{}0.08 (0.013)}\tabularnewline
 & {\small{}$\text{PPL}_{\infty}$} & {\small{}0.075 (0.024)} & {\small{}0.293 (0.043)} & {\small{}0.146 (0.021)} & {\small{}0.175 (0.032)} & \textbf{\small{}0.143 (0.03)}\tabularnewline
\hline 
\end{tabular}
\end{table} 

The results are presented in Table \ref{tab:dataset1_t}, which summarizes the simulation results for Dataset1 with the Student $t$ distribution error. With $t$ distribution, the performance of BEMP-nonlinear exhibits a sensitive result to the outliers induced from heavy-tailed error, and the relative performance compared to BEMP-poly worsens. Consequently, in some cases, BEMP-all exhibits inferior performance to BEMP-poly, which uses true polynomial structures for $h_2$. However, it still demonstrates the effect of using multiple proxies, with BEMP-all outperforming BEMP-nonlinear significantly. 

\begin{table}[htbp]
\centering
\caption{Monte Carlo means (standard errors) of MSE, $\text{PPL}_{1}$, and $\text{PPL}_{\infty}$ for homogeneous Dataset1 with Student $t$ distributed error}
\label{tab:dataset1_t}
\begin{tabular}{c|c|ccccc}
\hline 
\multicolumn{1}{c}{{\small{}Quantile}} & \multicolumn{1}{c}{} & {\small{}woME} & {\small{}Structural} & {\small{}BEMP-poly} & {\small{}BEMP-nonlinear} & {\small{}BEMP-all}\tabularnewline
\hline 
\multirow{3}{*}{{\small{}$p=0.1$}}
  & {\small{}MSE} & {\small{}0.123 (0.076)} & {\small{}0.319 (0.147)} & \textbf{\small{}0.161 (0.271)} & {\small{}0.311 (0.202)} & {\small{}0.181 (0.11)}\tabularnewline
 & {\small{}$\text{PPL}_{1}$} & {\small{}0.192 (0.051)} & {\small{}0.719 (0.19)} & {\small{}0.232 (0.174)} & {\small{}0.259 (0.108)} & \textbf{\small{}0.197 (0.062)}\tabularnewline
 & {\small{}$\text{PPL}_{\infty}$} & {\small{}0.253 (0.087)} & {\small{}0.884 (0.218)} & {\small{}0.304 (0.307)} & {\small{}0.414 (0.208)} & \textbf{\small{}0.296 (0.114)}\tabularnewline
\hline 
\multirow{3}{*}{{\small{}$p=0.25$}}
  & {\small{}MSE} & {\small{}0.031 (0.017)} & {\small{}0.164 (0.042)} & {\small{}0.064 (0.027)} & {\small{}0.118 (0.057)} & \textbf{\small{}0.053 (0.023)}\tabularnewline
 & {\small{}$\text{PPL}_{1}$} & {\small{}0.072 (0.01)} & {\small{}0.23 (0.047)} & {\small{}0.089 (0.023)} & {\small{}0.117 (0.031)} & \textbf{\small{}0.083 (0.017)}\tabularnewline
 & {\small{}$\text{PPL}_{\infty}$} & {\small{}0.086 (0.018)} & {\small{}0.313 (0.059)} & {\small{}0.121 (0.035)} & {\small{}0.175 (0.058)} & \textbf{\small{}0.112 (0.027)}\tabularnewline
\hline 
\multirow{3}{*}{{\small{}$p=0.5$}}
  & {\small{}MSE} & {\small{}0.019 (0.009)} & {\small{}0.128 (0.029)} & {\small{}0.039 (0.014)} & {\small{}0.083 (0.036)} & \textbf{\small{}0.033 (0.016)}\tabularnewline
 & {\small{}$\text{PPL}_{1}$} & {\small{}0.042 (0.005)} & {\small{}0.152 (0.026)} & {\small{}0.058 (0.008)} & {\small{}0.077 (0.016)} & \textbf{\small{}0.052 (0.008)}\tabularnewline
 & {\small{}$\text{PPL}_{\infty}$} & {\small{}0.051 (0.009)} & {\small{}0.222 (0.033)} & {\small{}0.077 (0.014)} & {\small{}0.12 (0.033)} & \textbf{\small{}0.07 (0.015)}\tabularnewline
\hline 
\multirow{3}{*}{{\small{}$p=0.75$}}
  & {\small{}MSE} & {\small{}0.035 (0.023)} & {\small{}0.172 (0.048)} & \textbf{\small{}0.089 (0.037)} & {\small{}0.159 (0.035)} & {\small{}0.098 (0.051)}\tabularnewline
 & {\small{}$\text{PPL}_{1}$} & {\small{}0.056 (0.012)} & {\small{}0.215 (0.036)} & {\small{}0.077 (0.017)} & {\small{}0.093 (0.014)} & \textbf{\small{}0.076 (0.019)}\tabularnewline
 & {\small{}$\text{PPL}_{\infty}$} & {\small{}0.074 (0.023)} & {\small{}0.299 (0.052)} & {\small{}0.124 (0.033)} & {\small{}0.174 (0.03)} & \textbf{\small{}0.123 (0.044)}\tabularnewline
\hline 
\multirow{3}{*}{{\small{}$p=0.9$}}
  & {\small{}MSE} & {\small{}0.118 (0.08)} & {\small{}0.321 (0.229)} & {\small{}0.17 (0.117)} & {\small{}0.213 (0.176)} & \textbf{\small{}0.164 (0.092)}\tabularnewline
 & {\small{}$\text{PPL}_{1}$} & {\small{}0.12 (0.045)} & {\small{}0.752 (0.237)} & {\small{}0.12 (0.116)} & {\small{}0.139 (0.103)} & \textbf{\small{}0.108 (0.056)}\tabularnewline
 & {\small{}$\text{PPL}_{\infty}$} & {\small{}0.178 (0.082)} & {\small{}0.917 (0.311)} & {\small{}0.207 (0.172)} & {\small{}0.246 (0.19)} & \textbf{\small{}0.189 (0.101)}\tabularnewline
\hline 
\end{tabular}
\end{table} 

The results with the gamma-distributed error in Table \ref{tab:dataset1_gamma} also present a similar trend. BEMP-all, using all available proxies, exhibits the most optimal performance, revealing that the proposed method has an efficient estimator with more information gained from combining proxy variables. However, when some information from the proxies is difficult to estimate, performance deteriorates moderately. We present the visualization of the gamma distribution error in Figure \ref{fig:fitted_res}, where the proposed method exhibits the largest performance gap between the woMEs. Figure \ref{fig:fitted_res} (a) reveals that the BEMP-nonlinear approach fails to capture the true relationship’s overall shape compared with BEMP-quad and worsens BEMP-all’s performance. However, BEMP-nonlinear still works better than the structural method, and the performance drop in BEMP-all is moderate, revealing a fitted line like that of BEMP-quad. For the extreme quantiles, BEMP-all and BEMP-quad fail to capture the curvature on the right side of the domain in $p=0.1$, whereas the methods in $p=0.9$ fail to capture the curvature on the left side of the domain.

\begin{table}[!htbp]
\centering
\caption{Monte Carlo means (standard errors) of MSE, $\text{PPL}_{1}$, and $\text{PPL}_{\infty}$ for homogeneous Dataset1 with gamma distributed error}
\label{tab:dataset1_gamma}
\begin{tabular}{c|c|ccccc}
\hline 
\multicolumn{1}{c}{{\small{}Quantile}} & \multicolumn{1}{c}{} & {\small{}woME} & {\small{}Structural} & {\small{}BEMP-poly} & {\small{}BEMP-nonlinear} & {\small{}BEMP-all}\tabularnewline
\hline 
\multirow{3}{*}{{\small{}$p=0.1$}}
  & {\small{}MSE} & {\small{}0.038 (0.024)} & {\small{}0.167 (0.047)} & {\small{}0.126 (0.032)} & {\small{}0.144 (0.04)} & \textbf{\small{}0.125 (0.038)}\tabularnewline
 & {\small{}$\text{PPL}_{1}$} & {\small{}0.064 (0.011)} & {\small{}0.247 (0.047)} & {\small{}0.085 (0.014)} & {\small{}0.088 (0.017)} & \textbf{\small{}0.078 (0.013)}\tabularnewline
 & {\small{}$\text{PPL}_{\infty}$} & {\small{}0.083 (0.022)} & {\small{}0.323 (0.059)} & {\small{}0.148 (0.028)} & {\small{}0.162 (0.036)} & \textbf{\small{}0.141 (0.03)}\tabularnewline
\hline 
\multirow{3}{*}{{\small{}$p=0.25$}}
  & {\small{}MSE} & {\small{}0.032 (0.015)} & {\small{}0.159 (0.045)} & {\small{}0.065 (0.028)} & {\small{}0.103 (0.036)} & \textbf{\small{}0.061 (0.026)}\tabularnewline
 & {\small{}$\text{PPL}_{1}$} & {\small{}0.053 (0.008)} & {\small{}0.224 (0.037)} & {\small{}0.068 (0.011)} & {\small{}0.082 (0.023)} & \textbf{\small{}0.062 (0.012)}\tabularnewline
 & {\small{}$\text{PPL}_{\infty}$} & {\small{}0.068 (0.015)} & {\small{}0.3 (0.048)} & {\small{}0.099 (0.025)} & {\small{}0.136 (0.04)} & \textbf{\small{}0.093 (0.024)}\tabularnewline
\hline 
\multirow{3}{*}{{\small{}$p=0.5$}}
  & {\small{}MSE} & {\small{}0.045 (0.02)} & {\small{}0.176 (0.054)} & {\small{}0.058 (0.027)} & {\small{}0.152 (0.082)} & \textbf{\small{}0.056 (0.035)}\tabularnewline
 & {\small{}$\text{PPL}_{1}$} & {\small{}0.069 (0.013)} & {\small{}0.273 (0.044)} & {\small{}0.084 (0.017)} & {\small{}0.136 (0.048)} & \textbf{\small{}0.079 (0.029)}\tabularnewline
 & {\small{}$\text{PPL}_{\infty}$} & {\small{}0.092 (0.022)} & {\small{}0.368 (0.054)} & {\small{}0.114 (0.029)} & {\small{}0.206 (0.087)} & \textbf{\small{}0.108 (0.045)}\tabularnewline
\hline 
\multirow{3}{*}{{\small{}$p=0.75$}}
  & {\small{}MSE} & {\small{}0.093 (0.04)} & {\small{}0.252 (0.088)} & \textbf{\small{}0.119 (0.079)} & {\small{}0.327 (0.142)} & {\small{}0.148 (0.082)}\tabularnewline
 & {\small{}$\text{PPL}_{1}$} & {\small{}0.126 (0.029)} & {\small{}0.463 (0.072)} & {\small{}0.177 (0.068)} & {\small{}0.274 (0.076)} & \textbf{\small{}0.176 (0.06)}\tabularnewline
 & {\small{}$\text{PPL}_{\infty}$} & {\small{}0.174 (0.047)} & {\small{}0.597 (0.089)} & \textbf{\small{}0.238 (0.105)} & {\small{}0.438 (0.145)} & {\small{}0.251 (0.098)}\tabularnewline
\hline 
\multirow{3}{*}{{\small{}$p=0.9$}}
  & {\small{}MSE} & {\small{}0.205 (0.101)} & {\small{}0.428 (0.177)} & \textbf{\small{}0.256 (0.182)} & {\small{}0.472 (0.178)} & {\small{}0.344 (0.176)}\tabularnewline
 & {\small{}$\text{PPL}_{1}$} & {\small{}0.254 (0.068)} & {\small{}0.951 (0.171)} & \textbf{\small{}0.318 (0.13)} & {\small{}0.401 (0.102)} & {\small{}0.327 (0.102)}\tabularnewline
 & {\small{}$\text{PPL}_{\infty}$} & {\small{}0.361 (0.116)} & {\small{}1.176 (0.2)} & \textbf{\small{}0.454 (0.217)} & {\small{}0.633 (0.189)} & {\small{}0.501 (0.188)}\tabularnewline
\hline 
\end{tabular}
\end{table} 

For Dataset2, the simulation results are summarized in Tables \ref{tab:dataset2_1}, \ref{tab:dataset2_2}, and \ref{tab:dataset2_3}. While for Dataset2, the heteroscedasticity and fluctuating function with large higher derivatives make the estimation difficult (Figure \ref{fig:fitted_res} (b)), and the overall result remains the same. In this dataset, the performance of BEMP-all outperforms that of all error distributions. Generalizing this limited simulation and suggesting that BEMP-all outperforms in all cases is difficult, but in practice, we rarely have information regarding the true polynomial structure of the proxy or the quality of each proxy; remarkably, the result is sufficiently encouraging to generally use the BEMP-all method. 


\begin{table}[!htbp]
\centering
\caption{\label{tab:dataset2_1} 
Monte Carlo means (standard errors) of MSE, $\text{PPL}_{1}$, and $\text{PPL}_{\infty}$ for heterogeneous Dataset2 with standard normal distributed error
}
\begin{tabular}{c|c|ccccc}
\hline
\multicolumn{1}{c}{} & \multicolumn{1}{c}{} & \multicolumn{5}{c}{Standard Normal Error}\tabularnewline
\hline 
\multicolumn{1}{c}{{\small{}Quantile}} & \multicolumn{1}{c}{} & {\small{}woME} & {\small{}Structural} & {\small{}BEMP-poly} & {\small{}BEMP-nonlinear} & {\small{}BEMP-all}\tabularnewline
\hline 
\multirow{3}{*}{{\small{}$p=0.1$}}
  & {\small{}MSE} & {\small{}0.048 (0.018)} & {\small{}0.481 (0.06)} & {\small{}0.333 (0.041)} & {\small{}0.429 (0.066)} & \textbf{\small{}0.225 (0.049)}\tabularnewline
 & {\small{}$\text{PPL}_{1}$} & {\small{}0.067 (0.008)} & {\small{}0.371 (0.054)} & {\small{}0.192 (0.02)} & {\small{}0.233 (0.032)} & \textbf{\small{}0.138 (0.023)}\tabularnewline
 & {\small{}$\text{PPL}_{\infty}$} & {\small{}0.089 (0.016)} & {\small{}0.616 (0.076)} & {\small{}0.359 (0.04)} & {\small{}0.449 (0.064)} & \textbf{\small{}0.252 (0.047)}\tabularnewline
\hline 
\multirow{3}{*}{{\small{}$p=0.25$}}
  & {\small{}MSE} & {\small{}0.022 (0.01)} & {\small{}0.32 (0.057)} & {\small{}0.204 (0.037)} & {\small{}0.296 (0.053)} & \textbf{\small{}0.112 (0.029)}\tabularnewline
 & {\small{}$\text{PPL}_{1}$} & {\small{}0.04 (0.006)} & {\small{}0.248 (0.046)} & {\small{}0.125 (0.018)} & {\small{}0.16 (0.026)} & \textbf{\small{}0.078 (0.014)}\tabularnewline
 & {\small{}$\text{PPL}_{\infty}$} & {\small{}0.05 (0.011)} & {\small{}0.412 (0.07)} & {\small{}0.227 (0.036)} & {\small{}0.308 (0.053)} & \textbf{\small{}0.135 (0.028)}\tabularnewline
\hline 
\multirow{3}{*}{{\small{}$p=0.5$}}
  & {\small{}MSE} & {\small{}0.018 (0.009)} & {\small{}0.158 (0.032)} & {\small{}0.094 (0.02)} & {\small{}0.166 (0.036)} & \textbf{\small{}0.063 (0.016)}\tabularnewline
 & {\small{}$\text{PPL}_{1}$} & {\small{}0.035 (0.005)} & {\small{}0.156 (0.033)} & {\small{}0.072 (0.009)} & {\small{}0.096 (0.018)} & \textbf{\small{}0.054 (0.008)}\tabularnewline
 & {\small{}$\text{PPL}_{\infty}$} & {\small{}0.043 (0.009)} & {\small{}0.233 (0.046)} & {\small{}0.119 (0.019)} & {\small{}0.18 (0.035)} & \textbf{\small{}0.085 (0.015)}\tabularnewline
\hline 
\multirow{3}{*}{{\small{}$p=0.75$}}
  & {\small{}MSE} & {\small{}0.021 (0.008)} & {\small{}0.221 (0.042)} & {\small{}0.112 (0.023)} & {\small{}0.2 (0.048)} & \textbf{\small{}0.07 (0.015)}\tabularnewline
 & {\small{}$\text{PPL}_{1}$} & {\small{}0.041 (0.005)} & {\small{}0.198 (0.038)} & {\small{}0.088 (0.012)} & {\small{}0.119 (0.023)} & \textbf{\small{}0.063 (0.008)}\tabularnewline
 & {\small{}$\text{PPL}_{\infty}$} & {\small{}0.051 (0.008)} & {\small{}0.303 (0.056)} & {\small{}0.144 (0.023)} & {\small{}0.22 (0.047)} & \textbf{\small{}0.098 (0.015)}\tabularnewline
\hline 
\multirow{3}{*}{{\small{}$p=0.9$}}
  & {\small{}MSE} & {\small{}0.033 (0.012)} & {\small{}0.488 (0.1)} & {\small{}0.282 (0.05)} & {\small{}0.437 (0.145)} & \textbf{\small{}0.155 (0.06)}\tabularnewline
 & {\small{}$\text{PPL}_{1}$} & {\small{}0.064 (0.009)} & {\small{}0.356 (0.071)} & {\small{}0.191 (0.027)} & {\small{}0.251 (0.07)} & \textbf{\small{}0.125 (0.032)}\tabularnewline
 & {\small{}$\text{PPL}_{\infty}$} & {\small{}0.082 (0.014)} & {\small{}0.608 (0.116)} & {\small{}0.331 (0.052)} & {\small{}0.466 (0.143)} & \textbf{\small{}0.205 (0.062)}\tabularnewline
\hline 
\end{tabular}
\end{table}

\begin{table}[!htbp]
\centering
\caption{\label{tab:dataset2_2} 
Monte Carlo means (standard errors) of MSE, $\text{PPL}_{1}$, and $\text{PPL}_{\infty}$ for heterogeneous Dataset2 with Student $t$ distributed error
}
\begin{tabular}{c|c|ccccc}
\hline
\multicolumn{1}{c}{} & \multicolumn{1}{c}{} & \multicolumn{5}{c}{Student $t$ Error}\tabularnewline
\hline 
\multicolumn{1}{c}{{\small{}Quantile}} & \multicolumn{1}{c}{} & {\small{}woME} & {\small{}Structural} & {\small{}BEMP-poly} & {\small{}BEMP-nonlinear} & {\small{}BEMP-all}\tabularnewline
\hline 
\multirow{3}{*}{{\small{}$p=0.1$}}
  & {\small{}MSE} & {\small{}0.116 (0.06)} & {\small{}0.553 (0.146)} & {\small{}0.323 (0.117)} & {\small{}0.39 (0.149)} & \textbf{\small{}0.285 (0.145)}\tabularnewline
 & {\small{}$\text{PPL}_{1}$} & {\small{}0.131 (0.035)} & {\small{}0.62 (0.188)} & {\small{}0.198 (0.057)} & {\small{}0.222 (0.072)} & \textbf{\small{}0.182 (0.071)}\tabularnewline
 & {\small{}$\text{PPL}_{\infty}$} & {\small{}0.194 (0.063)} & {\small{}0.887 (0.241)} & {\small{}0.358 (0.116)} & {\small{}0.418 (0.146)} & \textbf{\small{}0.326 (0.143)}\tabularnewline
\hline 
\multirow{3}{*}{{\small{}$p=0.25$}}
  & {\small{}MSE} & {\small{}0.035 (0.016)} & {\small{}0.339 (0.064)} & {\small{}0.222 (0.048)} & {\small{}0.296 (0.075)} & \textbf{\small{}0.137 (0.043)}\tabularnewline
 & {\small{}$\text{PPL}_{1}$} & {\small{}0.055 (0.008)} & {\small{}0.308 (0.052)} & {\small{}0.14 (0.024)} & {\small{}0.167 (0.036)} & \textbf{\small{}0.095 (0.023)}\tabularnewline
 & {\small{}$\text{PPL}_{\infty}$} & {\small{}0.071 (0.015)} & {\small{}0.492 (0.072)} & {\small{}0.252 (0.047)} & {\small{}0.312 (0.073)} & \textbf{\small{}0.162 (0.044)}\tabularnewline
\hline 
\multirow{3}{*}{{\small{}$p=0.5$}}
  & {\small{}MSE} & {\small{}0.022 (0.01)} & {\small{}0.161 (0.042)} & {\small{}0.096 (0.021)} & {\small{}0.167 (0.037)} & \textbf{\small{}0.066 (0.019)}\tabularnewline
 & {\small{}$\text{PPL}_{1}$} & {\small{}0.039 (0.006)} & {\small{}0.18 (0.039)} & {\small{}0.076 (0.01)} & {\small{}0.102 (0.018)} & \textbf{\small{}0.058 (0.009)}\tabularnewline
 & {\small{}$\text{PPL}_{\infty}$} & {\small{}0.05 (0.01)} & {\small{}0.263 (0.056)} & {\small{}0.125 (0.02)} & {\small{}0.187 (0.036)} & \textbf{\small{}0.09 (0.019)}\tabularnewline
\hline 
\multirow{3}{*}{{\small{}$p=0.75$}}
  & {\small{}MSE} & {\small{}0.036 (0.026)} & {\small{}0.251 (0.06)} & {\small{}0.132 (0.024)} & {\small{}0.183 (0.081)} & \textbf{\small{}0.086 (0.024)}\tabularnewline
 & {\small{}$\text{PPL}_{1}$} & {\small{}0.056 (0.013)} & {\small{}0.26 (0.05)} & {\small{}0.107 (0.013)} & {\small{}0.123 (0.038)} & \textbf{\small{}0.077 (0.014)}\tabularnewline
 & {\small{}$\text{PPL}_{\infty}$} & {\small{}0.075 (0.026)} & {\small{}0.377 (0.074)} & {\small{}0.174 (0.024)} & {\small{}0.214 (0.078)} & \textbf{\small{}0.12 (0.025)}\tabularnewline
\hline 
\multirow{3}{*}{{\small{}$p=0.9$}}
  & {\small{}MSE} & {\small{}0.097 (0.091)} & {\small{}0.544 (0.166)} & {\small{}0.308 (0.101)} & {\small{}0.341 (0.146)} & \textbf{\small{}0.22 (0.106)}\tabularnewline
 & {\small{}$\text{PPL}_{1}$} & {\small{}0.123 (0.051)} & {\small{}0.64 (0.163)} & {\small{}0.236 (0.06)} & {\small{}0.227 (0.073)} & \textbf{\small{}0.173 (0.058)}\tabularnewline
 & {\small{}$\text{PPL}_{\infty}$} & {\small{}0.167 (0.095)} & {\small{}0.938 (0.217)} & {\small{}0.386 (0.109)} & {\small{}0.394 (0.145)} & \textbf{\small{}0.28 (0.11)}\tabularnewline
\hline 
\end{tabular}
\end{table}

\bigskip
\begin{table}[!htbp]
\centering
\caption{\label{tab:dataset2_3} 
Monte Carlo means (standard errors) of MSE, $\text{PPL}_{1}$, and $\text{PPL}_{\infty}$ for heterogeneous Dataset2 with gamma distributed error
}
\begin{tabular}{c|c|ccccc}
\hline
\multicolumn{1}{c}{} & \multicolumn{1}{c}{} & \multicolumn{5}{c}{Gamma Error}\tabularnewline
\hline 
\multicolumn{1}{c}{{\small{}Quantile}} & \multicolumn{1}{c}{} & {\small{}woME} & {\small{}Structural} & {\small{}BEMP-poly} & {\small{}BEMP-nonlinear} & {\small{}BEMP-all}\tabularnewline
\hline 
\multirow{3}{*}{{\small{}$p=0.1$}}
  & {\small{}MSE} & {\small{}0.053 (0.022)} & {\small{}0.497 (0.11)} & {\small{}0.362 (0.059)} & {\small{}0.414 (0.106)} & \textbf{\small{}0.217 (0.051)}\tabularnewline
 & {\small{}$\text{PPL}_{1}$} & {\small{}0.071 (0.012)} & {\small{}0.446 (0.1)} & {\small{}0.215 (0.03)} & {\small{}0.226 (0.053)} & \textbf{\small{}0.138 (0.024)}\tabularnewline
 & {\small{}$\text{PPL}_{\infty}$} & {\small{}0.096 (0.022)} & {\small{}0.69 (0.14)} & {\small{}0.396 (0.059)} & {\small{}0.432 (0.106)} & \textbf{\small{}0.245 (0.049)}\tabularnewline
\hline 
\multirow{3}{*}{{\small{}$p=0.25$}}
  & {\small{}MSE} & {\small{}0.036 (0.017)} & {\small{}0.265 (0.075)} & {\small{}0.163 (0.036)} & {\small{}0.202 (0.072)} & \textbf{\small{}0.101 (0.03)}\tabularnewline
 & {\small{}$\text{PPL}_{1}$} & {\small{}0.056 (0.009)} & {\small{}0.29 (0.076)} & {\small{}0.114 (0.018)} & {\small{}0.122 (0.034)} & \textbf{\small{}0.08 (0.015)}\tabularnewline
 & {\small{}$\text{PPL}_{\infty}$} & {\small{}0.072 (0.016)} & {\small{}0.429 (0.103)} & {\small{}0.197 (0.036)} & {\small{}0.221 (0.07)} & \textbf{\small{}0.131 (0.029)}\tabularnewline
\hline 
\multirow{3}{*}{{\small{}$p=0.5$}}
  & {\small{}MSE} & {\small{}0.046 (0.023)} & {\small{}0.163 (0.056)} & {\small{}0.109 (0.03)} & {\small{}0.117 (0.056)} & \textbf{\small{}0.088 (0.033)}\tabularnewline
 & {\small{}$\text{PPL}_{1}$} & {\small{}0.062 (0.013)} & {\small{}0.268 (0.058)} & {\small{}0.093 (0.017)} & {\small{}0.093 (0.031)} & \textbf{\small{}0.077 (0.018)}\tabularnewline
 & {\small{}$\text{PPL}_{\infty}$} & {\small{}0.084 (0.024)} & {\small{}0.356 (0.068)} & {\small{}0.148 (0.031)} & {\small{}0.151 (0.059)} & \textbf{\small{}0.119 (0.034)}\tabularnewline
\hline 
\multirow{3}{*}{{\small{}$p=0.75$}}
  & {\small{}MSE} & {\small{}0.076 (0.042)} & {\small{}0.317 (0.091)} & {\small{}0.161 (0.06)} & {\small{}0.177 (0.085)} & \textbf{\small{}0.131 (0.059)}\tabularnewline
 & {\small{}$\text{PPL}_{1}$} & {\small{}0.092 (0.022)} & {\small{}0.465 (0.093)} & {\small{}0.14 (0.035)} & {\small{}0.141 (0.046)} & \textbf{\small{}0.114 (0.031)}\tabularnewline
 & {\small{}$\text{PPL}_{\infty}$} & {\small{}0.129 (0.042)} & {\small{}0.643 (0.113)} & {\small{}0.218 (0.065)} & {\small{}0.229 (0.087)} & \textbf{\small{}0.178 (0.06)}\tabularnewline
\hline 
\multirow{3}{*}{{\small{}$p=0.9$}}
  & {\small{}MSE} & {\small{}0.153 (0.062)} & {\small{}0.824 (0.209)} & {\small{}0.31 (0.108)} & {\small{}0.4 (0.171)} & \textbf{\small{}0.26 (0.096)}\tabularnewline
 & {\small{}$\text{PPL}_{1}$} & {\small{}0.157 (0.038)} & {\small{}1.12 (0.261)} & {\small{}0.247 (0.063)} & {\small{}0.26 (0.088)} & \textbf{\small{}0.188 (0.053)}\tabularnewline
 & {\small{}$\text{PPL}_{\infty}$} & {\small{}0.227 (0.067)} & {\small{}1.522 (0.318)} & {\small{}0.399 (0.116)} & {\small{}0.459 (0.173)} & \textbf{\small{}0.318 (0.1)}\tabularnewline
\hline 
\end{tabular}
\end{table}

\subsection{Estimation of Nonlinear Proxy}
To empirically examine the proposed model’s performance in estimating the proxy relationship and further validate the assumption regarding the effect of multiple proxies, we assess the model estimation of $h_3(x)$, the nonlinear relationship between the unobserved covariate and proxy. We evaluate the same metrics used in Section \ref{simulation_result} for the posterior result of $h_3(x)$. 

\begin{table}[htbp]
\centering
\caption{Monte Carlo means (standard errors) of MSE, $\text{PPL}_{1}$, and $\text{PPL}_{\infty}$ evaluated for $h_3$ in Dataset1 (top) and Dataset2 (bottom) 
}\label{tab:nonlin_sum}
\resizebox{\columnwidth}{!}{%
\begin{tabular}{c|c|cccccc}
\hline 
\multicolumn{1}{c}{} & \multicolumn{1}{c}{} & \multicolumn{2}{c}{Normal} & \multicolumn{2}{c}{Student $t$} & \multicolumn{2}{c}{Gamma}\tabularnewline
\hline 
\multicolumn{1}{c}{} & \multicolumn{1}{c}{} & \multicolumn{6}{c}{Dataset1}\tabularnewline
\hline 
\multicolumn{1}{c}{{\small{}Quantile}} & \multicolumn{1}{c}{} & BEMP-nonlinear & BEMP-all & BEMP-nonlinear & BEMP-all & BEMP-nonlinear & BEMP-all\tabularnewline
\hline 
\multirow{3}{*}{{\small{}$p=0.1$}} & {\small{}MSE} & \textbf{\small{}0.68 (0.375)} & {\small{}0.122 (0.081)} & {\small{}0.498 (0.173)} & \textbf{\small{}0.138 (0.08)} & {\small{}0.941 (0.323)} & \textbf{0.122 (0.076)}\tabularnewline
 & {\small{}$\text{PPL}_{1}$} & {\small{}0.43 (0.201)} & \textbf{\small{}0.17 (0.043)} & {\small{}0.322 (0.086)} & \textbf{\small{}0.175 (0.041)} & {\small{}0.539 (0.156)} & \textbf{0.168 (0.04)}\tabularnewline
 & {\small{}$\text{PPL}_{\infty}$} & {\small{}0.771 (0.387)} & \textbf{\small{}0.231 (0.082)} & {\small{}0.565 (0.172)} & \textbf{\small{}0.243 (0.08)} & {\small{}1 (0.316)} & \textbf{0.223 (0.076)}\tabularnewline
\hline 
\multirow{3}{*}{{\small{}$p=0.25$}} & {\small{}MSE} & {\small{}0.904 (0.489)} & \textbf{\small{}0.118 (0.074)} & {\small{}0.498 (0.217)} & \textbf{\small{}0.12 (0.084)} & {\small{}0.557 (0.249)} & \textbf{0.127 (0.07)}\tabularnewline
 & {\small{}$\text{PPL}_{1}$} & {\small{}0.534 (0.263)} & \textbf{\small{}0.156 (0.039)} & {\small{}0.332 (0.115)} & \textbf{\small{}0.163 (0.045)} & {\small{}0.372 (0.132)} & \textbf{0.16 (0.039)}\tabularnewline
 & {\small{}$\text{PPL}_{\infty}$} & {\small{}0.975 (0.505)} & \textbf{\small{}0.216 (0.075)} & {\small{}0.586 (0.222)} & \textbf{\small{}0.221 (0.086)} & {\small{}0.642 (0.255)} & \textbf{0.222 (0.072)}\tabularnewline
\hline 
\multirow{3}{*}{{\small{}$p=0.5$}} & {\small{}MSE} & {\small{}0.866 (0.364)} & \textbf{\small{}0.106 (0.06)} & {\small{}0.649 (0.489)} & \textbf{\small{}0.116 (0.075)} & {\small{}0.466 (0.178)} & \textbf{0.111 (0.084)}\tabularnewline
 & {\small{}$\text{PPL}_{1}$} & {\small{}0.514 (0.187)} & \textbf{\small{}0.157 (0.037)} & {\small{}0.409 (0.259)} & \textbf{\small{}0.155 (0.039)} & {\small{}0.306 (0.092)} & \textbf{0.157 (0.043)}\tabularnewline
 & {\small{}$\text{PPL}_{\infty}$} & {\small{}0.952 (0.368)} & \textbf{\small{}0.215 (0.065)} & {\small{}0.718 (0.503)} & \textbf{\small{}0.213 (0.075)} & {\small{}0.532 (0.18)} & \textbf{0.209 (0.083)}\tabularnewline
\hline 
\multirow{3}{*}{{\small{}$p=0.75$}} & {\small{}MSE} & {\small{}1.082 (0.308)} & \textbf{\small{}0.137 (0.102)} & {\small{}0.7 (0.298)} & \textbf{\small{}0.15 (0.095)} & {\small{}0.492 (0.166)} & \textbf{0.117 (0.067)}\tabularnewline
 & {\small{}$\text{PPL}_{1}$} & {\small{}0.611 (0.151)} & \textbf{\small{}0.168 (0.052)} & {\small{}0.417 (0.15)} & \textbf{\small{}0.176 (0.049)} & {\small{}0.323 (0.087)} & \textbf{0.157 (0.036)}\tabularnewline
 & {\small{}$\text{PPL}_{\infty}$} & {\small{}1.156 (0.303)} & \textbf{\small{}0.24 (0.102)} & {\small{}0.757 (0.298)} & \textbf{\small{}0.251 (0.096)} & {\small{}0.57 (0.169)} & \textbf{0.214 (0.069)}\tabularnewline
\hline 
\multirow{3}{*}{{\small{}$p=0.9$}} & {\small{}MSE} & {\small{}1.022 (0.308)} & \textbf{\small{}0.133 (0.099)} & {\small{}0.543 (0.191)} & \textbf{\small{}0.136 (0.092)} & {\small{}0.484 (0.25)} & \textbf{0.134 (0.08)}\tabularnewline
 & {\small{}$\text{PPL}_{1}$} & {\small{}0.576 (0.147)} & \textbf{\small{}0.163 (0.051)} & {\small{}0.349 (0.103)} & \textbf{\small{}0.167 (0.047)} & {\small{}0.318 (0.133)} & \textbf{0.173 (0.043)}\tabularnewline
 & {\small{}$\text{PPL}_{\infty}$} & {\small{}1.081 (0.3)} & \textbf{\small{}0.234 (0.099)} & {\small{}0.621 (0.198)} & \textbf{\small{}0.236 (0.091)} & {\small{}0.56 (0.257)} & \textbf{0.236 (0.081)}\tabularnewline
\hline 
\hline 
\multicolumn{1}{c}{} & \multicolumn{1}{c}{} & \multicolumn{6}{c}{Dataset2}\tabularnewline
\hline 
\multicolumn{1}{c}{{\small{}Quantile}} & \multicolumn{1}{c}{} & BEMP-nonlinear & BEMP-all & BEMP-nonlinear & BEMP-all & BEMP-nonlinear & BEMP-all\tabularnewline
\hline 
\multirow{3}{*}{{\small{}$p=0.1$}} & {\small{}MSE} & {\small{}1.143 (0.218)} & \textbf{\small{}0.14 (0.08)} & {\small{}0.878 (0.319)} & \textbf{\small{}0.135 (0.078)} & {\small{}1.168 (0.23)} & \textbf{0.117 (0.058)}\tabularnewline
 & {\small{}$\text{PPL}_{1}$} & {\small{}0.629 (0.109)} & \textbf{\small{}0.166 (0.042)} & {\small{}0.519 (0.154)} & \textbf{\small{}0.168 (0.042)} & {\small{}0.653 (0.113)} & \textbf{0.156 (0.03)}\tabularnewline
 & {\small{}$\text{PPL}_{\infty}$} & {\small{}1.203 (0.218)} & \textbf{\small{}0.236 (0.082)} & {\small{}0.948 (0.312)} & \textbf{\small{}0.235 (0.079)} & {\small{}1.23 (0.226)} & \textbf{0.213 (0.058)}\tabularnewline
\hline 
\multirow{3}{*}{{\small{}$p=0.25$}} & {\small{}MSE} & {\small{}1.153 (0.219)} & \textbf{\small{}0.111 (0.073)} & {\small{}1.111 (0.28)} & \textbf{\small{}0.132 (0.073)} & {\small{}1.066 (0.447)} & \textbf{0.126 (0.083)}\tabularnewline
 & {\small{}$\text{PPL}_{1}$} & {\small{}0.64 (0.115)} & \textbf{\small{}0.151 (0.042)} & {\small{}0.622 (0.142)} & \textbf{\small{}0.158 (0.038)} & {\small{}0.634 (0.243)} & \textbf{0.157 (0.042)}\tabularnewline
 & {\small{}$\text{PPL}_{\infty}$} & {\small{}1.214 (0.224)} & \textbf{\small{}0.206 (0.077)} & {\small{}1.185 (0.279)} & \textbf{\small{}0.221 (0.073)} & {\small{}1.162 (0.464)} & \textbf{0.221 (0.082)}\tabularnewline
\hline 
\multirow{3}{*}{{\small{}$p=0.5$}} & {\small{}MSE} & {\small{}1.149 (0.209)} & \textbf{\small{}0.124 (0.074)} & {\small{}1.119 (0.433)} & \textbf{\small{}0.123 (0.082)} & {\small{}0.916 (0.511)} & \textbf{0.119 (0.072)}\tabularnewline
 & {\small{}$\text{PPL}_{1}$} & {\small{}0.654 (0.105)} & \textbf{\small{}0.159 (0.038)} & {\small{}0.634 (0.25)} & \textbf{\small{}0.155 (0.042)} & {\small{}0.53 (0.277)} & \textbf{0.159 (0.037)}\tabularnewline
 & {\small{}$\text{PPL}_{\infty}$} & {\small{}1.212 (0.207)} & \textbf{\small{}0.221 (0.074)} & {\small{}1.198 (0.464)} & \textbf{\small{}0.218 (0.081)} & {\small{}0.977 (0.53)} & \textbf{0.211 (0.071)}\tabularnewline
\hline 
\multirow{3}{*}{{\small{}$p=0.75$}} & {\small{}MSE} & {\small{}1.123 (0.286)} & \textbf{\small{}0.125 (0.088)} & {\small{}1.05 (0.424)} & \textbf{\small{}0.124 (0.095)} & {\small{}0.71 (0.337)} & \textbf{0.116 (0.078)}\tabularnewline
 & {\small{}$\text{PPL}_{1}$} & {\small{}0.629 (0.165)} & \textbf{\small{}0.16 (0.047)} & {\small{}0.632 (0.212)} & \textbf{\small{}0.16 (0.053)} & {\small{}0.453 (0.188)} & \textbf{0.156 (0.043)}\tabularnewline
 & {\small{}$\text{PPL}_{\infty}$} & {\small{}1.181 (0.303)} & \textbf{\small{}0.219 (0.089)} & {\small{}1.153 (0.421)} & \textbf{\small{}0.215 (0.099)} & {\small{}0.8 (0.355)} & \textbf{0.218 (0.081)}\tabularnewline
\hline 
\multirow{3}{*}{{\small{}$p=0.9$}} & {\small{}MSE} & {\small{}1.12 (0.231)} & \textbf{\small{}0.146 (0.094)} & {\small{}0.824 (0.336)} & \textbf{\small{}0.129 (0.08)} & {\small{}0.722 (0.24)} & \textbf{0.127 (0.093)}\tabularnewline
 & {\small{}$\text{PPL}_{1}$} & {\small{}0.631 (0.122)} & \textbf{\small{}0.173 (0.047)} & {\small{}0.509 (0.176)} & \textbf{\small{}0.166 (0.042)} & {\small{}0.447 (0.124)} & \textbf{0.165 (0.048)}\tabularnewline
 & {\small{}$\text{PPL}_{\infty}$} & {\small{}1.19 (0.23)} & \textbf{\small{}0.243 (0.093)} & {\small{}0.904 (0.342)} & \textbf{\small{}0.23 (0.081)} & {\small{}0.812 (0.242)} & \textbf{0.226 (0.093)}\tabularnewline
\hline 
\end{tabular}
}
\end{table}

Table \ref{tab:nonlin_sum} summarizes the result for Dataset1 and Dataset2 for all types of error distribution. We verify that the proposed method successfully estimates the nonlinear relationship between the proxy and true covariate with moderate performance across all metrics. Additionally, BEMP-all exhibits superior performance to BEMP-nonlinear in estimating the proxy’s nonlinear relationship. This result supports our assumption that information from other proxies improves the estimation of the proxy’s current relationship. The improved estimation of the relationship between the proxy and covariate directly affects the unobserved covariate’s estimation, which might be important for the estimated quantile function’s performance. 

\subsection{Estimation of Unobserved Covariate}

Although the estimation of the regression function $g_p$ is the primary focus in most cases, the inference of the mismeasured covariate $x$ can be another important interest. To further investigate the proposed method, we examine the posterior samples to estimate the unobserved covariate $x$. Table \ref{tab:X_mse} presents the MSE between the posterior samples of the unobserved covariate $x$ and their true value in Dataset1 with a normal error. Notably, with the naive approach using $w_1$ instead of $x$, the MSE is $\sigma_1^2$ in expectation, which is $1$ in our simulation. The MSE of the estimation of $x$ exhibits patterns consistent with the other parameters. For all the models that we test, the MSE is smaller  than that of the naive approach . BEMP-nonlinear and structural methods exhibit similar performances, whereas adding a nonlinear proxy still boosts the estimation of $x$. This is evident in that BEMP-all generally outperforms BEMP-quad in all the quantiles, which, again, demonstrates multiple proxies’ effectiveness. The results from the other cases exhibit similar patterns (Appendix).

\begin{table}[htbp]
\centering
\caption{Monte Carlo means (standard errors) of MSE for estimation of unobserved covariate $x$ in Dataset1 with normal error}
\label{tab:X_mse}
\begin{tabular}{c|cccc}
\hline 
\multicolumn{1}{c}{{\small{}Quantile}} & {\small{}Structural} & {\small{}BEMP-poly} & {\small{}BEMP-nonlinear} & {\small{}BEMP-all}\tabularnewline
\hline 
{\small{}$p=0.1$} & 0.914 (0.043) & 0.259 (0.037) & 0.756 (0.087) & \textbf{0.229 (0.051)}\tabularnewline
{\small{}$p=0.25$} & 0.915 (0.043) & 0.25 (0.038) & 0.78 (0.095) & \textbf{0.221 (0.052)}\tabularnewline
{\small{}$p=0.5$} & 0.913 (0.045) & 0.248 (0.035) & 0.787 (0.1) & \textbf{0.219 (0.05)}\tabularnewline
{\small{}$p=0.75$} & 0.908 (0.046) & 0.254 (0.036) & 0.899 (0.121) & \textbf{0.219 (0.052)}\tabularnewline
{\small{}$p=0.9$} & 0.902 (0.046) & 0.261 (0.038) & 0.947 (0.128) & \textbf{0.215 (0.054)}\tabularnewline
\hline 
\end{tabular}
\end{table}

\section{Application to Administrative Data}\label{real_data}
We apply the proposed method to a real dataset that includes asset and income variables. Statistics Korea released microdata from the Survey of Household Finances and Living Conditions (SFLC), incorporating administrative data obtained from other government institutions. The released dataset includes basic demographic variables for 18,064 families and various economic features, such as salary income, property income, assets, asset management plans, debt, and debt repayment capacity for each family unit collected in 2020. 
	
This application aims to determine the quantile relationship between assets and true salary income. Income data provide critical information for a wide range of policies. However, administrative salary income is prone to measurement error, so the direct use of this information can precipitate misleading inferences \citep{moore2000income, davern2005effect}. We consider using administrative salary income and property income as two types of proxies: one is exposed to additive error, and the other is a correlated proxy. These values are suitable for use as proxies because it is reasonable to assume that property income and salary income have a high correlation \citep{lerman1985income}. Consequently, the model is described as follows:
\begin{align*}
	\text{asset}_{i}| \text{true salary income}_{i} & = g_p(\text{true salary income}_{i})+e_{i}\\
	\text{administrative salary income}_{i} & =\text{true salary income}_{i}+u_{1i}\\
	\text{property income}_{i} & =\alpha_{0}+\alpha_{1}\text{true salary income}_{i}+u_{2i}
\end{align*}
	
Notably, we assume that we do not  observe $\text{true salary income}_i$ but instead observe multiple proxies, $\text{administrative income}_i$ and $\text{property income}_i$. To model the correlated proxy $\text{property income}$, we utilize a linear regression for the relationship between the proxy and covariate with parameter $\boldsymbol{\alpha}$. As a preprocessing step, we eliminate the extreme quantiles (i.e., $0.999$ and $0.001$ percentiles in terms of each variable). After preprocessing, the data comprises 11,317 family units. Further, we attempt a log transform for asset variables to alleviate data skewness and improve model convergence. Following \cite{thompson2010bayesian}'s suggestion, we take $N=30$ knots equally spaced over the range of variables, which is log-transformed administrative salary income.

\begin{figure}[htbp]
\centering
\includegraphics[width=0.9\textwidth]{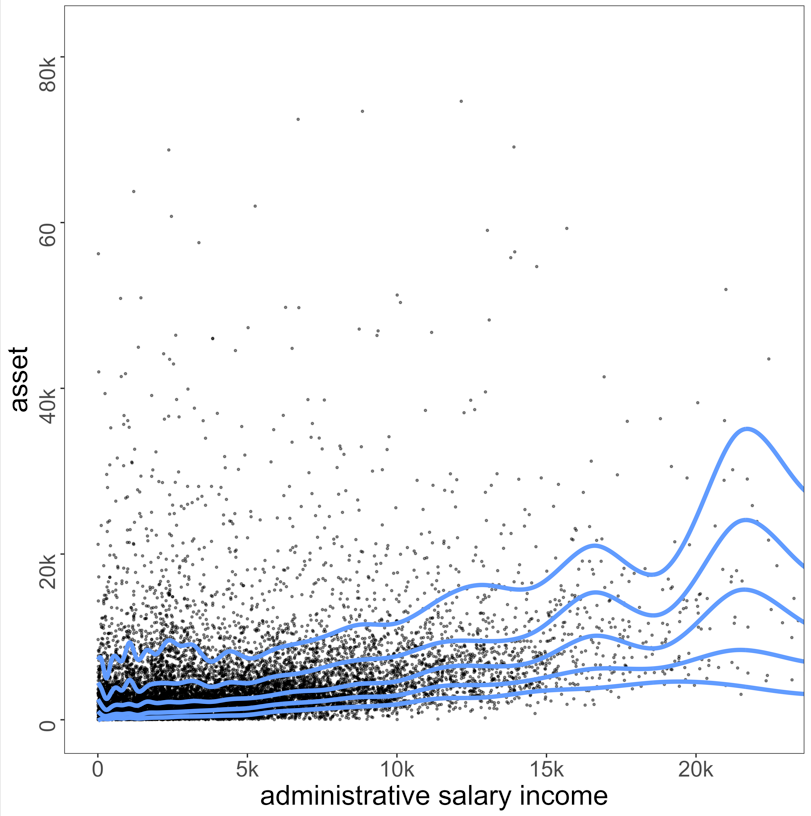} 
\caption{Fitted line for income data application. The $x$-axis denotes the administrative salary income from the survey but is suspected of exhibiting a survey measurement error. The $y$-axis represents the assets. The fitted line is based on the quantile $p\in\{0.1,0.25,0.5,0.75,0.9\}$. }\label{fig:income}
\end{figure}

We investigate the effect of $\text{true salary income}$ on $\text{asset}$ in different quantile $p\in\left\{0.1,\ 0.25,\ 0.5,\ 0.75,\ 0.9\right\}$. Figure \ref{fig:income} presents the resulting quantile lines. In Figure \ref{fig:income}, the fitted relation’s spread is not parallel, implying that a heterogeneous effect exists. The fitted quantile function of $p=0.9$ presents a larger gap between the other fitted quantile functions, which indicates that the conditional distribution of $\text{asset}$ is not symmetrical and right-skewed. More interestingly, for higher quantiles, such as $p=0.9$ and $p=0.75$, the fitted line in the lower level of administrative salary income ($\le 5\text{k}$) generally curved upward, with its highest point in $\text{administrative salary income} \simeq 3\text{k}$. 

This outcome might be attributable to various reasons. For example, people with their assets not structurally proportional to salary income  or those with the upper end of asset value might structurally misreport their salary income, which is in line with the literature \citep{moore2000income, stocke2006attitudes, valet2019comparing}. In either case, nonparametric quantile regression can derive a novel insight  that was impossible through parametric regression and is more informative in combining proxies than the naïve approach.
	
\section{Conclusion}\label{conclude}

This study proposes a Bayesian quantile regression estimation method that integrates multiple proxies obtained from multiple datasets. The proposed method has two notable advantages compared to previous methods. First, the proposed approach handles multiple data sources with different relationships to the same covariate, whereas previous methods were developed for handling a single proxy with only an additive error or limited structure. Another strength is that the proposed method is not restricted to linear regression functions and alleviates parametric assumptions regarding the regression function of interest such that the conditional distribution’s dispersion can be investigated more precisely. 

A simulation study on various datasets demonstrates that the proposed method can accommodate multiple proxies with linear and nonlinear relationships with the true covariate. We demonstrate that the proposed method is promising for capturing the underlying relationship, effective in incorporating multiple proxies simultaneously, and making reliable estimations of unobserved covariates and their relationships with proxies. Further, we presented an application of this methodology using a public SFLC dataset and provided the underlying relationship between assets and salary income in the presence of multiple income records.

This study has some limitations. We adopted a spline function for quantile regression; thus, the fitted regression line’s behavior tends to be erratic near the boundaries, which is a well-known characteristic of the spline approach \citep{friedman2001elements}. Further, noteworthily,  both natural cubic spline quantile regression \citep{thompson2010bayesian} and P-spline quantile regression \citep{lang2004bayesian} use the asymmetric Laplace distribution (ALD) for errors because Bayesian quantile regression has been widely studied by specifying the ALD \citep{yu2001bayesian}. We chose ALD in this study because it facilitates Bayesian inference and computation in the complicated additive models considered here; the result was reasonably flexible, as presented in Section \ref{simulation}. With a similar framework, alternative approaches such as \cite{dunson2005approximate, kottas2009bayesian, taddy2010bayesian}'s can also be applied to settings with multiple proxies. We leave the extension of this parametric assumption for future research. Finally, the current study relies on a benchmark variable’s presence, which is necessary for estimating the unobserved covariate’s scale. The alleviation of this assumption and its subsequent analysis, such as robustness, remain a topic for future research. 

\section*{Acknowledgements}
Ick Hoon Jin is partially supported by the Yonsei University Research Fund of 2019-22-0210 and by the National Research Foundation (NRF) Korea (NRF 2020R1A2C1A01009881). Jongho Im's research is also supported by the NRF Korea (NRF-2021R1C1C1014407).

\bibliography{reference}

\begin{thebibliography}{}

\bibitem[\protect\citeauthoryear{Aigner, Hsiao, Kapteyn, and Wansbeek}{Aigner
  et~al.}{1984}]{aigner1984latent}
Aigner, D.~J., C.~Hsiao, A.~Kapteyn, and T.~Wansbeek (1984).
\newblock Latent variable models in econometrics.
\newblock {\em Handbook of econometrics\/}~{\em 2}, 1321--1393.

\bibitem[\protect\citeauthoryear{Berg, Im, Zhu, Colin, and Li}{Berg
  et~al.}{2021}]{berg2021}
Berg, E., J.~Im, Z.~Zhu, L.-B. Colin, and J.~Li (2021).
\newblock Integration of statistical and administrative agricultural data from
  namibia.
\newblock {\em Statistical Journal of the IAOS\/}~{\em 37}, 557--578.

\bibitem[\protect\citeauthoryear{Berry, Carroll, and Ruppert}{Berry
  et~al.}{2002}]{berry2002bayesian}
Berry, S.~M., R.~J. Carroll, and D.~Ruppert (2002).
\newblock Bayesian smoothing and regression splines for measurement error
  problems.
\newblock {\em Journal of the American Statistical Association\/}~{\em
  97\/}(457), 160--169.

\bibitem[\protect\citeauthoryear{Brezger and Lang}{Brezger and
  Lang}{2006}]{brezger2006generalized}
Brezger, A. and S.~Lang (2006).
\newblock Generalized structured additive regression based on bayesian
  p-splines.
\newblock {\em Computational Statistics \& Data Analysis\/}~{\em 50\/}(4),
  967--991.

\bibitem[\protect\citeauthoryear{Brown}{Brown}{1982}]{brown1982robust}
Brown, M.~L. (1982).
\newblock Robust line estimation with errors in both variables.
\newblock {\em Journal of the American Statistical Association\/}~{\em
  77\/}(377), 71--79.

\bibitem[\protect\citeauthoryear{Carroll, Maca, and Ruppert}{Carroll
  et~al.}{1999}]{carroll1999nonparametric}
Carroll, R.~J., J.~D. Maca, and D.~Ruppert (1999).
\newblock Nonparametric regression in the presence of measurement error.
\newblock {\em Biometrika\/}~{\em 86\/}(3), 541--554.

\bibitem[\protect\citeauthoryear{Carroll, Ruppert, Stefanski, and
  Crainiceanu}{Carroll et~al.}{2006}]{carroll2006measurement}
Carroll, R.~J., D.~Ruppert, L.~A. Stefanski, and C.~M. Crainiceanu (2006).
\newblock {\em Measurement error in nonlinear models: a modern perspective}.
\newblock CRC press.

\bibitem[\protect\citeauthoryear{Clayton}{Clayton}{1992}]{clayton1992}
Clayton, D.~G. (1992).
\newblock {\em Models for the Analysis of Cohort and Case-Control Studies with
  Inaccurately Measured Exposures.}
\newblock In Statistical Models for Longitudinal Studies of Exposure and
  Health, edited by James H. Dwyer, Manning Feinleib, Peter Lippert, and Hans
  Hoffmeister, 301–31. New York: Oxford University Press.

\bibitem[\protect\citeauthoryear{Davern, Rodin, Beebe, and Call}{Davern
  et~al.}{2005}]{davern2005effect}
Davern, M., H.~Rodin, T.~J. Beebe, and K.~T. Call (2005).
\newblock The effect of income question design in health surveys on family
  income, poverty and eligibility estimates.
\newblock {\em Health services research\/}~{\em 40\/}(5p1), 1534--1552.

\bibitem[\protect\citeauthoryear{Delaigle, Hall, and Meister}{Delaigle
  et~al.}{2008}]{delaigle2008deconvolution}
Delaigle, A., P.~Hall, and A.~Meister (2008).
\newblock On deconvolution with repeated measurements.
\newblock {\em The Annals of Statistics\/}~{\em 36\/}(2), 665--685.

\bibitem[\protect\citeauthoryear{Dunson and Taylor}{Dunson and
  Taylor}{2005}]{dunson2005approximate}
Dunson, D.~B. and J.~A. Taylor (2005).
\newblock Approximate bayesian inference for quantiles.
\newblock {\em Journal of Nonparametric Statistics\/}~{\em 17\/}(3), 385--400.

\bibitem[\protect\citeauthoryear{Eliers and Marx}{Eliers and
  Marx}{1996}]{eliers1996flexible}
Eliers, P. and B.~Marx (1996).
\newblock Flexible smoothing using b-splines and penalized likelihood (with
  comments and rejoinder).
\newblock {\em Statistical Science\/}~{\em 11}, 1200--1224.

\bibitem[\protect\citeauthoryear{Eubank}{Eubank}{1999}]{eubank1999nonparametric}
Eubank, R.~L. (1999).
\newblock {\em Nonparametric regression and spline smoothing}.
\newblock CRC press.

\bibitem[\protect\citeauthoryear{Fan}{Fan}{1992}]{fan1992design}
Fan, J. (1992).
\newblock Design-adaptive nonparametric regression.
\newblock {\em Journal of the American statistical Association\/}~{\em
  87\/}(420), 998--1004.

\bibitem[\protect\citeauthoryear{Fenske, Kneib, and Hothorn}{Fenske
  et~al.}{2011}]{fenske2011identifying}
Fenske, N., T.~Kneib, and T.~Hothorn (2011).
\newblock Identifying risk factors for severe childhood malnutrition by
  boosting additive quantile regression.
\newblock {\em Journal of the American Statistical Association\/}~{\em
  106\/}(494), 494--510.

\bibitem[\protect\citeauthoryear{Filmer and Pritchett}{Filmer and
  Pritchett}{2001}]{filmer2001estimating}
Filmer, D. and L.~H. Pritchett (2001).
\newblock Estimating wealth effects without expenditure data—or tears: an
  application to educational enrollments in states of india.
\newblock {\em Demography\/}~{\em 38\/}(1), 115--132.

\bibitem[\protect\citeauthoryear{Firpo, Galvao, and Song}{Firpo
  et~al.}{2017}]{firpo2017measurement}
Firpo, S., A.~F. Galvao, and S.~Song (2017).
\newblock Measurement errors in quantile regression models.
\newblock {\em Journal of econometrics\/}~{\em 198\/}(1), 146--164.

\bibitem[\protect\citeauthoryear{Friedman, Hastie, Tibshirani, et~al.}{Friedman
  et~al.}{2001}]{friedman2001elements}
Friedman, J., T.~Hastie, R.~Tibshirani, et~al. (2001).
\newblock {\em The elements of statistical learning}, Volume~1.
\newblock Springer series in statistics New York.

\bibitem[\protect\citeauthoryear{Fuller}{Fuller}{1987}]{fuller2006measurement}
Fuller, W.~A. (1987).
\newblock {\em Measurement error models}.
\newblock John Wiley \& Sons.

\bibitem[\protect\citeauthoryear{Fuller}{Fuller}{2009}]{fuller2009sampling}
Fuller, W.~A. (2009).
\newblock {\em Sampling Statistics}.
\newblock John Wiley \& Sons.

\bibitem[\protect\citeauthoryear{Gamerman and Lopes}{Gamerman and
  Lopes}{2006}]{gamerman2006markov}
Gamerman, D. and H.~F. Lopes (2006).
\newblock {\em Markov chain Monte Carlo: stochastic simulation for Bayesian
  inference}.
\newblock CRC Press.

\bibitem[\protect\citeauthoryear{Gelfand and Ghosh}{Gelfand and
  Ghosh}{1998}]{gelfand1998model}
Gelfand, A.~E. and S.~K. Ghosh (1998).
\newblock Model choice: a minimum posterior predictive loss approach.
\newblock {\em Biometrika\/}~{\em 85\/}(1), 1--11.

\bibitem[\protect\citeauthoryear{Green and Silverman}{Green and
  Silverman}{1993}]{green1993nonparametric}
Green, P.~J. and B.~W. Silverman (1993).
\newblock {\em Nonparametric regression and generalized linear models: a
  roughness penalty approach}.
\newblock Crc Press.

\bibitem[\protect\citeauthoryear{H{\"a}rdle}{H{\"a}rdle}{1986}]{hardle1986approximations}
H{\"a}rdle, W. (1986).
\newblock Approximations to the mean integrated squared error with applications
  to optimal bandwidth selection for nonparametric regression function
  estimators.
\newblock {\em Journal of multivariate analysis\/}~{\em 18\/}(1), 150--168.

\bibitem[\protect\citeauthoryear{Hausman, Liu, Luo, and Palmer}{Hausman
  et~al.}{2021}]{hausman2021errors}
Hausman, J., H.~Liu, Y.~Luo, and C.~Palmer (2021).
\newblock Errors in the dependent variable of quantile regression models.
\newblock {\em Econometrica\/}~{\em 89\/}(2), 849--873.

\bibitem[\protect\citeauthoryear{He and Liang}{He and
  Liang}{2000}]{he2000quantile}
He, X. and H.~Liang (2000).
\newblock Quantile regression estimates for a class of linear and partially
  linear errors-in-variables models.
\newblock {\em Statistica Sinica\/}~{\em 10\/}(1), 129--140.

\bibitem[\protect\citeauthoryear{Hoerl and Kennard}{Hoerl and
  Kennard}{1970}]{hoerl1970ridge}
Hoerl, A.~E. and R.~W. Kennard (1970).
\newblock Ridge regression: Biased estimation for nonorthogonal problems.
\newblock {\em Technometrics\/}~{\em 12\/}(1), 55--67.

\bibitem[\protect\citeauthoryear{Koenker and {Bassett}}{Koenker and
  {Bassett}}{1978}]{koenker1978regression}
Koenker, R. and J.~G. {Bassett} (1978).
\newblock Regression quantiles.
\newblock {\em Econometrica: journal of the Econometric Society\/}~{\em
  46\/}(1), 33--50.

\bibitem[\protect\citeauthoryear{Kottas and Krnjaji{\'c}}{Kottas and
  Krnjaji{\'c}}{2009}]{kottas2009bayesian}
Kottas, A. and M.~Krnjaji{\'c} (2009).
\newblock Bayesian semiparametric modelling in quantile regression.
\newblock {\em Scandinavian Journal of Statistics\/}~{\em 36\/}(2), 297--319.

\bibitem[\protect\citeauthoryear{Kozumi and Kobayashi}{Kozumi and
  Kobayashi}{2011}]{kozumi2011gibbs}
Kozumi, H. and G.~Kobayashi (2011).
\newblock Gibbs sampling methods for bayesian quantile regression.
\newblock {\em Journal of statistical computation and simulation\/}~{\em
  81\/}(11), 1565--1578.

\bibitem[\protect\citeauthoryear{Lang and Brezger}{Lang and
  Brezger}{2004}]{lang2004bayesian}
Lang, S. and A.~Brezger (2004).
\newblock Bayesian p-splines.
\newblock {\em Journal of computational and graphical statistics\/}~{\em
  13\/}(1), 183--212.

\bibitem[\protect\citeauthoryear{Lerman and Yitzhaki}{Lerman and
  Yitzhaki}{1985}]{lerman1985income}
Lerman, R.~I. and S.~Yitzhaki (1985).
\newblock Income inequality effects by income source: A new approach and
  applications to the united states.
\newblock {\em The review of economics and statistics\/}~{\em 67}, 151--156.

\bibitem[\protect\citeauthoryear{Leulescu and Agafitei}{Leulescu and
  Agafitei}{2013}]{leulescu13}
Leulescu, A. and M.~Agafitei (2013).
\newblock {\em Statistical matching: a model based approach for data
  integration}.
\newblock Eurostat methodologies and working paper.

\bibitem[\protect\citeauthoryear{Li and Vuong}{Li and
  Vuong}{1998}]{li1998nonparametric}
Li, T. and Q.~Vuong (1998).
\newblock Nonparametric estimation of the measurement error model using
  multiple indicators.
\newblock {\em Journal of Multivariate Analysis\/}~{\em 65\/}(2), 139--165.

\bibitem[\protect\citeauthoryear{Lubotsky and Wittenberg}{Lubotsky and
  Wittenberg}{2006}]{lubotsky2006interpretation}
Lubotsky, D. and M.~Wittenberg (2006).
\newblock Interpretation of regressions with multiple proxies.
\newblock {\em The Review of Economics and Statistics\/}~{\em 88\/}(3),
  549--562.

\bibitem[\protect\citeauthoryear{Mazumder}{Mazumder}{2001}]{mazumder2001earnings}
Mazumder, B. (2001).
\newblock Earnings mobility in the us: A new look at intergenerational
  inequality.
\newblock Technical report, Federal Reserve Bank of Chicago.

\bibitem[\protect\citeauthoryear{Montes-Rojas et~al.}{Montes-Rojas
  et~al.}{2011}]{montes2011quantile}
Montes-Rojas, G. et~al. (2011).
\newblock Quantile regression with classical additive measurement errors.
\newblock {\em Economics Bulletin\/}~{\em 31\/}(4), 2863--2868.

\bibitem[\protect\citeauthoryear{Moore, Stinson, and Welniak}{Moore
  et~al.}{2000}]{moore2000income}
Moore, J.~C., L.~L. Stinson, and E.~J. Welniak (2000).
\newblock Income measurement error in surveys: A review.
\newblock {\em Journal of Official Statistics-Stockholm-\/}~{\em 16\/}(4),
  331--362.

\bibitem[\protect\citeauthoryear{Richardson and Gilks}{Richardson and
  Gilks}{1993}]{richardson1993}
Richardson, S. and W.~R. Gilks (1993).
\newblock A bayesian approach to measurement error problems in epidemiology
  using conditional independence models.
\newblock {\em American Journal of Epidemiology\/}~{\em 138\/}(6), 430--442.

\bibitem[\protect\citeauthoryear{Ruppert}{Ruppert}{2002}]{ruppert2002selecting}
Ruppert, D. (2002).
\newblock Selecting the number of knots for penalized splines.
\newblock {\em Journal of computational and graphical statistics\/}~{\em
  11\/}(4), 735--757.

\bibitem[\protect\citeauthoryear{Schennach}{Schennach}{2007}]{schennach2007instrumental}
Schennach, S.~M. (2007).
\newblock Instrumental variable estimation of nonlinear errors-in-variables
  models.
\newblock {\em Econometrica\/}~{\em 75\/}(1), 201--239.

\bibitem[\protect\citeauthoryear{Schennach}{Schennach}{2008}]{schennach2008quantile}
Schennach, S.~M. (2008).
\newblock Quantile regression with mismeasured covariates.
\newblock {\em Econometric Theory\/}~{\em 24\/}(4), 1010--1043.

\bibitem[\protect\citeauthoryear{Solon}{Solon}{1992}]{solon1992intergenerational}
Solon, G. (1992).
\newblock Intergenerational income mobility in the united states.
\newblock {\em The American Economic Review\/}~{\em 82\/}(3), 393--408.

\bibitem[\protect\citeauthoryear{Stock{\'e}}{Stock{\'e}}{2006}]{stocke2006attitudes}
Stock{\'e}, V. (2006).
\newblock Attitudes toward surveys, attitude accessibility and the effect on
  respondents’ susceptibility to nonresponse.
\newblock {\em Quality and Quantity\/}~{\em 40\/}(2), 259--288.

\bibitem[\protect\citeauthoryear{Taddy and Kottas}{Taddy and
  Kottas}{2010}]{taddy2010bayesian}
Taddy, M.~A. and A.~Kottas (2010).
\newblock A bayesian nonparametric approach to inference for quantile
  regression.
\newblock {\em Journal of Business \& Economic Statistics\/}~{\em 28\/}(3),
  357--369.

\bibitem[\protect\citeauthoryear{Thompson, Cai, Moyeed, Reeve, and
  Stander}{Thompson et~al.}{2010}]{thompson2010bayesian}
Thompson, P., Y.~Cai, R.~Moyeed, D.~Reeve, and J.~Stander (2010).
\newblock Bayesian nonparametric quantile regression using splines.
\newblock {\em Computational {S}tatistics \& {D}ata {A}nalysis\/}~{\em
  54\/}(4), 1138--1150.

\bibitem[\protect\citeauthoryear{Tibshirani}{Tibshirani}{1996}]{tibshirani1996regression}
Tibshirani, R. (1996).
\newblock Regression shrinkage and selection via the lasso.
\newblock {\em Journal of the Royal Statistical Society: Series B
  (Methodological)\/}~{\em 58\/}(1), 267--288.

\bibitem[\protect\citeauthoryear{Valet, Adriaans, and Liebig}{Valet
  et~al.}{2019}]{valet2019comparing}
Valet, P., J.~Adriaans, and S.~Liebig (2019).
\newblock Comparing survey data and administrative records on gross earnings:
  nonreporting, misreporting, interviewer presence and earnings inequality.
\newblock {\em Quality \& Quantity\/}~{\em 53\/}(1), 471--491.

\bibitem[\protect\citeauthoryear{Waldmann, Kneib, Yue, Lang, and
  Flexeder}{Waldmann et~al.}{2013}]{waldmann2013bayesian}
Waldmann, E., T.~Kneib, Y.~R. Yue, S.~Lang, and C.~Flexeder (2013).
\newblock Bayesian semiparametric additive quantile regression.
\newblock {\em Statistical Modelling\/}~{\em 13\/}(3), 223--252.

\bibitem[\protect\citeauthoryear{Wang, Stefanski, and Zhu}{Wang
  et~al.}{2012}]{wang2012corrected}
Wang, H.~J., L.~A. Stefanski, and Z.~Zhu (2012).
\newblock Corrected-loss estimation for quantile regression with covariate
  measurement errors.
\newblock {\em Biometrika\/}~{\em 99\/}(2), 405--421.

\bibitem[\protect\citeauthoryear{Wei and Carroll}{Wei and
  Carroll}{2009}]{wei2009quantile}
Wei, Y. and R.~J. Carroll (2009).
\newblock Quantile regression with measurement error.
\newblock {\em Journal of the American Statistical Association\/}~{\em
  104\/}(487), 1129--1143.

\bibitem[\protect\citeauthoryear{Yu, Lu, and Stander}{Yu
  et~al.}{2003}]{yu2003quantile}
Yu, K., Z.~Lu, and J.~Stander (2003).
\newblock Quantile regression: applications and current research areas.
\newblock {\em Journal of the Royal Statistical Society: Series D (The
  Statistician)\/}~{\em 52\/}(3), 331--350.

\bibitem[\protect\citeauthoryear{Yu and Moyeed}{Yu and
  Moyeed}{2001}]{yu2001bayesian}
Yu, K. and R.~A. Moyeed (2001).
\newblock Bayesian quantile regression.
\newblock {\em Statistics \& Probability Letters\/}~{\em 54\/}(4), 437--447.

\bibitem[\protect\citeauthoryear{Yue and Rue}{Yue and
  Rue}{2011}]{yue2011bayesian}
Yue, Y.~R. and H.~Rue (2011).
\newblock Bayesian inference for additive mixed quantile regression models.
\newblock {\em Computational Statistics \& Data Analysis\/}~{\em 55\/}(1),
  84--96.

\bibitem[\protect\citeauthoryear{Zimmerman}{Zimmerman}{1992}]{zimmerman1992regression}
Zimmerman, D.~J. (1992).
\newblock Regression toward mediocrity in economic stature.
\newblock {\em The American Economic Review\/}~{\em 82\/}(3), 409--429.

\end{thebibliography}
\clearpage 

\section*{Appendix: Additional Simulation Results}
\renewcommand{\arraystretch}{0.781}
\begin{table}[htbp]
    \centering
    \caption{\small{}Monte Carlo means (standard errors) of MSE for naive model which directly treats the proxy with error $w_2$ as true covariate and MSE for BEMP with P-spline evaluated in Dataset1 (top) and Dataset2 (bottom)
    }\label{tab:BSQR_MSE_performance}
    \resizebox{!}{\dimexpr\height-2\baselineskip\relax}{%
    \begin{tabular}{c|c|cccc}
    \hline 
    \multicolumn{1}{c}{} & \multicolumn{1}{c}{} & \multicolumn{4}{c}{{\small{}Dataset1}}\tabularnewline
    \hline 
    \multicolumn{1}{c}{{\small{}Quantile}} & \multicolumn{1}{c}{} & {\small{}Naive} & {\small{}BEMP-poly} & {\small{}BEMP-nonlinear} & {\small{}BEMP-all}\tabularnewline
    \hline 
    \multirow{3}{*}{{\small{}$p=0.1$}} & {\small{}Normal} & {\small{}0.622 (0.09)} & {\small{}0.309 (0.035)} & {\small{}0.226 (0.05)} & \textbf{\small{}0.175 (0.044)}\tabularnewline
     & {\small{}Student $t$} & {\small{}0.614 (0.117)} & {\small{}0.302 (0.071)} & {\small{}0.257 (0.076)} & \textbf{\small{}0.21 (0.066)}\tabularnewline
     & {\small{}Gamma} & {\small{}0.622 (0.124)} & {\small{}0.385 (0.065)} & {\small{}0.233 (0.069)} & \textbf{\small{}0.143 (0.042)}\tabularnewline
    \hline 
    \multirow{3}{*}{{\small{}$p=0.25$}} & {\small{}Normal} & {\small{}0.401 (0.058)} & {\small{}0.209 (0.036)} & {\small{}0.155 (0.036)} & \textbf{\small{}0.11 (0.03)}\tabularnewline
     & {\small{}Student $t$} & {\small{}0.444 (0.077)} & {\small{}0.246 (0.043)} & {\small{}0.195 (0.039)} & \textbf{\small{}0.135 (0.034)}\tabularnewline
     & {\small{}Gamma} & {\small{}0.375 (0.082)} & {\small{}0.208 (0.042)} & {\small{}0.11 (0.04)} & \textbf{\small{}0.095 (0.025)}\tabularnewline
    \hline 
    \multirow{3}{*}{{\small{}$p=0.5$}} & {\small{}Normal} & {\small{}0.257 (0.08)} & {\small{}0.114 (0.021)} & {\small{}0.101 (0.027)} & \textbf{\small{}0.069 (0.023)}\tabularnewline
     & {\small{}Student $t$} & {\small{}0.287 (0.081)} & {\small{}0.134 (0.02)} & {\small{}0.11 (0.025)} & \textbf{\small{}0.078 (0.02)}\tabularnewline
     & {\small{}Gamma} & {\small{}0.262 (0.068)} & {\small{}0.174 (0.035)} & \textbf{\small{}0.072 (0.026)} & {\small{}0.099 (0.033)}\tabularnewline
    \hline 
    \multirow{3}{*}{{\small{}$p=0.75$}} & {\small{}Normal} & {\small{}0.316 (0.094)} & {\small{}0.152 (0.022)} & {\small{}0.092 (0.029)} & \textbf{\small{}0.074 (0.018)}\tabularnewline
     & {\small{}Student $t$} & {\small{}0.37 (0.08)} & {\small{}0.184 (0.024)} & {\small{}0.111 (0.029)} & \textbf{\small{}0.1 (0.023)}\tabularnewline
     & {\small{}Gamma} & {\small{}0.403 (0.091)} & {\small{}0.293 (0.056)} & {\small{}0.101 (0.039)} & \textbf{\small{}0.167 (0.052)}\tabularnewline
    \hline 
    \multirow{3}{*}{{\small{}$p=0.9$}} & {\small{}Normal} & {\small{}0.715 (0.162)} & {\small{}0.258 (0.039)} & {\small{}0.125 (0.034)} & \textbf{\small{}0.11 (0.026)}\tabularnewline
     & {\small{}Student $t$} & {\small{}0.727 (0.156)} & {\small{}0.347 (0.077)} & \textbf{\small{}0.165 (0.044)} & {\small{}0.174 (0.068)}\tabularnewline
     & {\small{}Gamma} & {\small{}0.94 (0.212)} & {\small{}0.486 (0.093)} & \textbf{\small{}0.199 (0.086)} & {\small{}0.266 (0.085)}\tabularnewline
    \hline 
    \hline 
    \multicolumn{1}{c}{} & \multicolumn{1}{c}{} & \multicolumn{4}{c}{{\small{}Dataset2}}\tabularnewline
    \hline 
    \multicolumn{1}{c}{{\small{}Quantile}} & \multicolumn{1}{c}{} & {\small{}Naive} & {\small{}BEMP-poly} & {\small{}BEMP-nonlinear} & {\small{}BEMP-all}\tabularnewline
    \hline 
    \multirow{3}{*}{{\small{}$p=0.1$}} & {\small{}Normal} & {\small{}0.231 (0.062)} & {\small{}0.083 (0.021)} & {\small{}0.096 (0.027)} & \textbf{\small{}0.06 (0.022)}\tabularnewline
     & {\small{}Student $t$} & {\small{}0.292 (0.099)} & {\small{}0.135 (0.047)} & {\small{}0.134 (0.046)} & \textbf{\small{}0.127 (0.04)}\tabularnewline
     & {\small{}Gamma} & {\small{}0.176 (0.045)} & {\small{}0.084 (0.023)} & {\small{}0.095 (0.024)} & \textbf{\small{}0.072 (0.026)}\tabularnewline
    \hline 
    \multirow{3}{*}{{\small{}$p=0.25$}} & {\small{}Normal} & {\small{}0.154 (0.044)} & {\small{}0.068 (0.015)} & {\small{}0.079 (0.024)} & \textbf{\small{}0.048 (0.016)}\tabularnewline
     & {\small{}Student $t$} & {\small{}0.195 (0.057)} & {\small{}0.084 (0.02)} & {\small{}0.097 (0.023)} & \textbf{\small{}0.06 (0.019)}\tabularnewline
     & {\small{}Gamma} & {\small{}0.144 (0.05)} & {\small{}0.082 (0.02)} & {\small{}0.089 (0.024)} & \textbf{\small{}0.069 (0.02)}\tabularnewline
    \hline 
    \multirow{3}{*}{{\small{}$p=0.5$}} & {\small{}Normal} & {\small{}0.147 (0.037)} & {\small{}0.056 (0.014)} & {\small{}0.071 (0.029)} & \textbf{\small{}0.043 (0.014)}\tabularnewline
     & {\small{}Student $t$} & {\small{}0.161 (0.038)} & {\small{}0.06 (0.011)} & {\small{}0.075 (0.023)} & \textbf{\small{}0.046 (0.013)}\tabularnewline
     & {\small{}Gamma} & {\small{}0.183 (0.05)} & {\small{}0.088 (0.027)} & {\small{}0.093 (0.028)} & \textbf{\small{}0.075 (0.029)}\tabularnewline
    \hline 
    \multirow{3}{*}{{\small{}$p=0.75$}} & {\small{}Normal} & {\small{}0.213 (0.028)} & {\small{}0.063 (0.014)} & {\small{}0.078 (0.019)} & \textbf{\small{}0.048 (0.015)}\tabularnewline
     & {\small{}Student $t$} & {\small{}0.206 (0.027)} & {\small{}0.08 (0.019)} & {\small{}0.087 (0.027)} & \textbf{\small{}0.062 (0.021)}\tabularnewline
     & {\small{}Gamma} & {\small{}0.234 (0.092)} & {\small{}0.124 (0.037)} & \textbf{\small{}0.113 (0.037)} & {\small{}0.115 (0.04)}\tabularnewline
    \hline 
    \multirow{3}{*}{{\small{}$p=0.9$}} & {\small{}Normal} & {\small{}0.171 (0.04)} & {\small{}0.076 (0.018)} & {\small{}0.085 (0.025)} & \textbf{\small{}0.056 (0.02)}\tabularnewline
     & {\small{}Student $t$} & {\small{}0.211 (0.084)} & {\small{}0.127 (0.051)} & {\small{}0.127 (0.043)} & \textbf{\small{}0.127 (0.053)}\tabularnewline
     & {\small{}Gamma} & {\small{}0.349 (0.116)} & {\small{}0.175 (0.074)} & \textbf{\small{}0.171 (0.067)} & {\small{}0.18 (0.081)}\tabularnewline
    \hline 
    \end{tabular}{\small\par}
    }
    \par
    \end{table}

\begin{table}[htbp]
\centering
\caption{Monte Carlo means (standard errors) of MSE for estimation of unobserved covariate $x$ in Dataset1 with Student $t$ error, Gamma error.}
\label{tab:X_mse_app1}
\begin{tabular}{c|cccc}
\hline 
\multicolumn{1}{c}{} & \multicolumn{4}{c}{Student $t$ Error}\tabularnewline
\hline 
\multicolumn{1}{c}{{\small{}Quantile}} & {\small{}Structural} & {\small{}BEMP-poly} & {\small{}BEMP-nonlinear} & {\small{}BEMP-all}\tabularnewline
\hline 
{\small{}$p=0.1$} & 0.913 (0.043) & 0.266 (0.035) & 0.722 (0.067) & \textbf{0.219 (0.053)}\tabularnewline
{\small{}$p=0.25$} & 0.902 (0.043) & 0.258 (0.037) & 0.722 (0.069) & \textbf{0.223 (0.049)}\tabularnewline
{\small{}$p=0.5$} & 0.902 (0.044) & 0.253 (0.039) & 0.723 (0.086) & \textbf{0.218 (0.05)}\tabularnewline
{\small{}$p=0.75$} & 0.9 (0.043) & 0.255 (0.037) & 0.726 (0.093) & \textbf{0.217 (0.047)}\tabularnewline
{\small{}$p=0.9$} & 0.9 (0.044) & 0.262 (0.034) & 0.721 (0.072) & \textbf{0.224 (0.058)}\tabularnewline
\hline 
\hline 
\multicolumn{1}{c}{} & \multicolumn{4}{c}{Gamma Error}\tabularnewline
\hline 
\multicolumn{1}{c}{{\small{}Quantile}} & {\small{}Structural} & {\small{}BEMP-poly} & {\small{}BEMP-nonlinear} & {\small{}BEMP-all}\tabularnewline
\hline 
{\small{}$p=0.1$} & 0.914 (0.044) & 0.262 (0.035) & 0.781 (0.122) & \textbf{0.225 (0.053)}\tabularnewline
{\small{}$p=0.25$} & 0.904 (0.043) & 0.254 (0.038) & 0.714 (0.072) & \textbf{0.211 (0.051)}\tabularnewline
{\small{}$p=0.5$} & 0.902 (0.044) & 0.258 (0.04) & 0.697 (0.061) & \textbf{0.211 (0.052)}\tabularnewline
{\small{}$p=0.75$} & 0.901 (0.044) & 0.26 (0.038) & 0.721 (0.069) & \textbf{0.219 (0.05)}\tabularnewline
{\small{}$p=0.9$} & 0.901 (0.044) & 0.269 (0.039) & 0.716 (0.056) & \textbf{0.229 (0.054)}\tabularnewline
\hline 
\end{tabular}
\end{table} 

\begin{table}[htbp]
\centering
\caption{Monte Carlo means (standard errors) of MSE for estimation of unobserved covariate $x$ in Dataset2 with Normal error, Student $t$ error, and Gamma error.}
\label{tab:X_mse_app2}
\begin{tabular}{c|cccc}
\hline 
\multicolumn{1}{c}{} & \multicolumn{4}{c}{Standard Normal Error}\tabularnewline
\hline 
\multicolumn{1}{c}{{\small{}Quantile}} & {\small{}Structural} & {\small{}BEMP-poly} & {\small{}BEMP-nonlinear} & {\small{}BEMP-all}\tabularnewline
\hline 
{\small{}$p=0.1$} & 0.913 (0.045) & 0.25 (0.036) & 0.969 (0.053) & \textbf{0.22 (0.057)}\tabularnewline
{\small{}$p=0.25$} & 0.913 (0.044) & 0.224 (0.034) & 0.97 (0.039) & \textbf{0.215 (0.048)}\tabularnewline
{\small{}$p=0.5$} & 0.914 (0.044) & 0.213 (0.031) & 0.969 (0.046) & \textbf{0.201 (0.046)}\tabularnewline
{\small{}$p=0.75$} & 0.913 (0.043) & 0.23 (0.032) & 0.957 (0.088) & \textbf{0.204 (0.046)}\tabularnewline
{\small{}$p=0.9$} & 0.913 (0.043) & 0.242 (0.034) & 0.936 (0.097) & \textbf{0.219 (0.046)}\tabularnewline
\hline 
\hline 
\multicolumn{1}{c}{} & \multicolumn{4}{c}{Student $t$ Error}\tabularnewline
\hline 
\multicolumn{1}{c}{{\small{}Quantile}} & {\small{}Structural} & {\small{}BEMP-poly} & {\small{}BEMP-nonlinear} & {\small{}BEMP-all}\tabularnewline
\hline 
{\small{}$p=0.1$} & 0.913 (0.044) & 0.258 (0.039) & 0.793 (0.11) & \textbf{0.231 (0.051)}\tabularnewline
{\small{}$p=0.25$} & 0.913 (0.044) & 0.238 (0.032) & 0.931 (0.108) & \textbf{0.217 (0.055)}\tabularnewline
{\small{}$p=0.5$} & 0.913 (0.043) & 0.226 (0.033) & 0.949 (0.085) & \textbf{0.207 (0.045)}\tabularnewline
{\small{}$p=0.75$} & 0.913 (0.043) & 0.235 (0.03) & 0.856 (0.102) & \textbf{0.204 (0.047)}\tabularnewline
{\small{}$p=0.9$} & 0.913 (0.044) & 0.241 (0.032) & 0.764 (0.091) & \textbf{0.219 (0.051)}\tabularnewline
\hline 
\hline 
\multicolumn{1}{c}{} & \multicolumn{4}{c}{Gamma Error}\tabularnewline
\hline 
\multicolumn{1}{c}{{\small{}Quantile}} & {\small{}Structural} & {\small{}BEMP-poly} & {\small{}BEMP-nonlinear} & {\small{}BEMP-all}\tabularnewline
\hline 
{\small{}$p=0.1$} & 0.913 (0.041) & 0.22 (0.035) & 0.968 (0.077) & \textbf{0.184 (0.049)}\tabularnewline
{\small{}$p=0.25$} & 0.913 (0.044) & 0.194 (0.029) & 0.823 (0.12) & \textbf{0.171 (0.041)}\tabularnewline
{\small{}$p=0.5$} & 0.908 (0.046) & 0.187 (0.027) & 0.727 (0.097) & \textbf{0.167 (0.041)}\tabularnewline
{\small{}$p=0.75$} & 0.908 (0.046) & 0.196 (0.027) & 0.687 (0.085) & \textbf{0.174 (0.041)}\tabularnewline
{\small{}$p=0.9$} & 0.901 (0.044) & 0.218 (0.03) & 0.72 (0.088) & \textbf{0.188 (0.047)}\tabularnewline
\hline 
\end{tabular}
\end{table}

\end{document}